\documentclass[aps,pra,twocolumn,10pt,showpacs,preprintnumbers,amsmath,amssymb,floatfix,longbibliography]{revtex4-2}
\usepackage{graphicx}
\usepackage{bm}
\usepackage{color}
\usepackage{mathrsfs}
\usepackage{soul}
\usepackage{gensymb}
\usepackage{textcomp}

\newcommand{\bra}[1]{\langle #1|}
\newcommand{\ket}[1]{|#1\rangle}

\newcommand{\affA}{\vspace{1ex}$^{1}$Institute for Theoretical Physics, Institute of Physics, University of Amsterdam, Science Park 904, 1098 XH Amsterdam, The Netherlands}
\newcommand{\affB}{$^{2}$QuSoft, Science Park 123, 1098 XG Amsterdam, The Netherlands}
\newcommand{\affC}{$^{3}$Department of Physics, Indian Institute of Technology Madras, Chennai 600036, India}
\newcommand{\affD}{$^{4}$Center for Quantum Information, Communication and Computing, Indian Institute of Technology Madras, Chennai 600036, India}
\newcommand{\affE}{$^{5}$School of Physics, University of Sydney, NSW 2006 Sydney, Australia}

\newcommand{\affF}{$^{6}$ARC Centre of Excellence for Engineered Quantum Systems, University of Sydney, NSW 2006, Australia}
\newcommand{\affG}{$^{7}$Sydney Nano Institute, University of Sydney, NSW 2006, Australia}

\usepackage[normalem]{ulem}
\usepackage{physics}
\usepackage{multirow}

\usepackage{makecell}

\usepackage{hyperref}
\usepackage{xcolor}
\hypersetup{
    colorlinks,
    linkcolor={red!70!black},
    citecolor={green!65!black},
    urlcolor={blue!80!black}
}

\begin{document}

\title{Multilevel Electromagnetically Induced Transparency Cooling}

\author{
  Katya Fouka$^{1,2}$}\email{s.k.fouka@uva.nl}
\author{Athreya Shankar$^{3,4}$}
\author{Ting Rei Tan$^{5,6,7}$}
\author{Arghavan Safavi-Naini$^{1,2}$
}

\affiliation{\affA \\ \affB \\ \affC\\ \affD \\ \affE \\ \affF \\ \affG}

\date{\today}

\begin{abstract}

Electromagnetically Induced Transparency (EIT) cooling is a well-established method for preparing trapped ion systems in their motional ground state. However, isolating a three-level system, as required for EIT cooling, is often challenging or impractical. Nonetheless, multilevel systems can inherently host dark states. In this work, we extend the EIT cooling framework to such multilevel systems. We develop a formalism to accurately determine the cooling rate in the weak sideband coupling regime and provide an approximate estimate for cooling rates beyond this regime, without the need for explicit simulation of the motional degree of freedom. We clarify the connection between the cooling rate and the absorption spectrum, offering a pathway for efficient near-ground-state cooling of ions with complex electronic structures.
\end{abstract}

\maketitle

\section{INTRODUCTION}

Trapped ions have played an important role in advancing quantum simulation, computation, sensing, and metrology \cite{Leibfried2003,RevModPhys.87.637,RevModPhys.93.025001}. These applications typically require the ability to manipulate the motional degrees of freedom of the ions. For example, cooling the ion crystal to its motional ground state is necessary for high gate fidelities~\cite{PhysRevLett.117.060505,PhysRevLett.117.060504} as well as enabling quantum enhanced metrology and sensing, which often requires preparation of multi-partite entangled states \cite{metrology, sensing}. In particular, phonon mediated interactions between ions, arising from the coupling of the atomic states with the collective motional degrees of freedom of the ion crystal, are crucial in the operation of logical gates and state preparation protocols \cite{PhysRevLett.74.4091,PhysRevLett.82.1971, PhysRevLett.82.1835,Leibfried2003}. However, large ion crystals, required for useful quantum operations, present significant challenges. As the crystal size increases, the corresponding growth of motional modes demands cooling over a broad bandwidth to counteract environmental heating and ensure high-fidelity operations \cite{PhysRevLett.81.1525, PhysRevLett.117.060504,PhysRevLett.117.140501}.

Experiments with trapped ions typically begin with a Doppler cooling stage \cite{PhysRevA.20.1521}, followed by sub-Doppler cooling techniques to overcome the temperature limitation imposed by the linewidth of the addressed transition \cite{Itano_1995}. Commonly used ground-state cooling techniques include Sisyphus cooling \cite{Dalibard:89,G_Birkl_1994,Joshi_2020, PhysRevLett.119.043001}, resolved sideband cooling \cite{PhysRevLett.62.403,PhysRevLett.75.4011,PhysRevLett.83.4713} and electromagnetically induced transparency (EIT) cooling \cite{Morigi2000, eit_exp,Morigi_eit}. Although Sisyphus cooling is capable of simultaneously cooling a broad spectrum of motional modes, its cooling limit is not sufficiently low \cite{Dalibard:89,G_Birkl_1994,Joshi_2020, PhysRevLett.119.043001}. On the other hand, resolved sideband cooling provides a low cooling limit. However, its narrow cooling bandwidth results in an overall cooling duration that increases with the number of ions \cite{PhysRevA.102.043110,PhysRevResearch.5.023022}. In contrast, EIT cooling offers both fast near ground state preparation and a broad cooling bandwidth, positioning it as one of the most attractive sub-Doppler cooling techniques \cite{PhysRevLett.110.153002,PhysRevA.93.053401,PhysRevA.98.023424,PhysRevA.99.023409, PhysRevLett.122.053603,PhysRevLett.125.053001,PhysRevLett.125.053001,Qiao2021,PhysRevApplied.18.014022}.

Although \textit{single} EIT cooling \cite{Morigi2000, eit_exp,Morigi_eit} is predominantly linked to a closed three-level system, its applicability has been successfully extended to more complex configurations. In particular, EIT cooling has been demonstrated in closed four-level systems, referred to as \textit{double} EIT cooling \cite{double_eit_theory1,double_eit_theory2,Semerikov2018,PhysRevLett.125.053001,Qiao2021}. Additionally, it has been recently performed in multilevel systems by engineering an open three-level system using additional laser sources beyond those required for cooling \cite{PhysRevLett.133.113204} as well as in an open four-level system \cite{wu2024}. 

The core principle of EIT cooling relies on using the dark states of the light-atom system. Multilevel systems can naturally support dark states as long as the coherent population trapping condition is met \cite{Multilevel_dark}. This suggests that EIT cooling can be performed without the necessity of isolating a three-level system, whether open or closed.

In this paper, we numerically investigate \textit{multilevel} EIT cooling in systems hosting dark states. We begin, in Sec.~II., by briefly recalling the principles of single EIT cooling. We summarize established techniques for estimating cooling rates while introducing an alternative method. Furthermore, we assess the validity of these techniques within the sideband coupling regimes. In Sec.~III., we model the multilevel cooling transition of a realistic ion for arbitrary nuclear spin $I$ and study multilevel EIT cooling while testing the cooling estimation techniques for $I=1$ and $I=7$.

\section{SINGLE ELECTROMAGNETICALLY INDUCED TRANSPARENCY COOLING}

\subsection{Background}

\begin{figure}[htbp]
\includegraphics[scale=1.0]{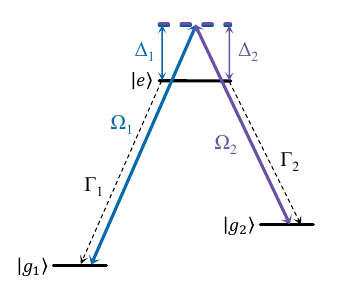}
\centering
\caption{Level scheme of single electromagnetically induced transparency cooling: Solid arrows connecting the $\ket{g_{j}} \rightarrow \ket{e}$ represent the two lasers with Rabi frequencies $\Omega_j$, detuned by $\Delta_j$. Dashed arrows represent decay from $\ket{e} \rightarrow \ket{g_j}$ with decay rate $\Gamma_j$.}
\label{eit_levels}
\end{figure}

We start by reviewing single electromagnetically induced transparency (EIT) cooling as introduced in \cite{Morigi2000,Morigi_eit, eit_exp}. This scheme utilizes a closed three-level system comprised of an excited state $\ket{e}$ coupled to two long-lived ground states $\ket{g_1}$, $\ket{g_2}$ (see Fig.~\ref{eit_levels}) in a harmonic potential. The system is driven by two counter-propagating lasers: a `drive' laser that addresses the 
$\ket{g_1} \rightarrow \ket{e}$ transition with Rabi frequency $\Omega_1$ and detuning $\Delta_1$, and a `probe' laser that excites the $\ket{g_2} \rightarrow \ket{e}$ with Rabi frequency $\Omega_2$ and detuning $\Delta_2$. If $\Delta_1=\Delta_2$, the steady state of the system is a so-called dark state with no excited state population. In this section, we use $\Delta_d\equiv\Delta_1=\Delta_2$ to indicate that the dark state condition is satisfied. 

In the Lamb-Dicke (LD) regime, cooling can be achieved by canceling the resonant carrier transition $\ket{n} \rightarrow \ket{n}$ while simultaneously inducing an imbalance between the off-resonant red- and blue-sideband transitions, which correspond to $\ket{n} \rightarrow \ket{n-1}$  and $\ket{n} \rightarrow \ket{n+1}$ respectively, so that the removal of phonons is favoured.

The optimal cooling scheme for a given system will result in near ground-state cooling in the shortest time possible. To determine the conditions for optimal cooling we first need to consider the dynamics of a typical trapped three-level ion. We model the trap along the $x$ direction as a harmonic potential with frequency $\omega_0$. The Hamiltonian in a rotating frame is
\begin{flalign}\label{ham_m}
    H &= \omega_0 a^{\dagger}a + \Delta_1 \sigma_{g_1,g_1} + \Delta_2 \sigma_{g_2,g_2} \nonumber \\
    & -\frac{\Omega_1}{2} e^{ik_1 x}\sigma_{e,g_1} - \frac{\Omega_2}{2}e^{ik_2 x}\sigma_{e,g_2} + \text{h.c}.
\end{flalign}
where $a \,(a^\dagger)$ is the annihilation (creation) operator,  $\Delta_{j}$ are the detunings of the  $\ket{g_{j}} \rightarrow \ket{e}$ transitions, $\sigma_{\mu,\nu} = \ket{\mu}\bra{\nu}$ and $k_j$ are the wave numbers of the lasers ($j = 1,2$). Expanding the Hamiltonian up to first order in the LD regime, we obtain, 
\begin{equation} \label{ld_eq}
    H \approx \omega_0 a^{\dagger}a + H_R + V_1(a + a^{\dagger}), \\
\end{equation}
with
\begin{flalign}
    H_R &= \Delta_1 \sigma_{g_1,g_1} + \Delta_2 \sigma_{g_2,g_2} \nonumber \\
    & -\frac{\Omega_1}{2} \sigma_{e,g_1} - \frac{\Omega_2}{2}\sigma_{e,g_2} + \text{h.c}. \label{restion_Ham} \\
    V_1 &= \frac{i\eta\Omega_1}{2}\sigma_{e,g_{1}} -\frac{i\eta\Omega_2}{2}\sigma_{e,g_{2}} + \text{h.c.}
\end{flalign}
and  $\eta_2=-\eta_1=\eta$, where $\eta$ is the LD parameter defined as $\eta = k\sqrt{\hbar/2m\omega_0}$. 

Finally, we include the spontaneous emission decay process from  $\ket{e} \rightarrow \ket{g_j}$ at rate $\Gamma_j$ and arrive at the Lindblad master equation, 
\begin{equation}\label{full_dynamics}
    \dot{\rho} = -i\left[ H, \rho\right] + \mathcal{L}\rho 
\end{equation}
where 
\begin{equation}\label{eq:diss}
    \mathcal{L}\rho = \sum_{j}\Gamma_j \left[\sigma_{g_j,e}\rho\sigma_{e,g_j}-\frac{1}{2}\{\sigma_{e,e},\rho\}\right]. 
\end{equation}
We note that Eq.\eqref{eq:diss} is incomplete, as it disregards the recoil of the ion due to spontaneous emission \cite{Cirac_Zoller_cooling,Molmer:93}. However, this term scales with $\eta^2$ and depends on the excited state population. Since we restrict our analysis to $\eta \leq 0.1$ and we eliminate the carrier transition, the recoil's influence on the dynamics is negligible and it can be safely omitted.

\subsection{Cooling optimization}

The optimal parameters are dependent on the details of the system under consideration. Our aim is to determine these parameters without explicitly calculating the full dynamics, which includes both the electronic and motional degrees of freedom. In this section, we take three routes for finding the optimal cooling parameters and we evaluate their predictions against numerical simulations of the full dynamics Eq.\eqref{full_dynamics}. The numerical cooling rate $W_{\text{exp}}$ is obtained by fitting the calculated mean phonon number $\langle n \rangle$ to

\begin{equation}\label{exp_decay}
\langle n \rangle = \langle n \rangle_0\,\text{exp}[-W_{\text{exp}}t]+\langle n \rangle_f\,\left(1-\text{exp}[-W_{\text{exp}}t]\right),
\end{equation}
where $\langle n \rangle_{0}$ and $\langle n \rangle_{f}$ is the initial and final mean phonon number respectively.

\subsubsection{Rest ion}

\begin{figure}[htpb]
\includegraphics[width=\linewidth]{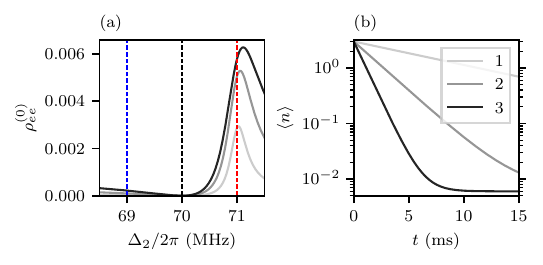}
\centering
\caption{(a) Absorption profile of the rest ion and  (b) corresponding cooling dynamics for  (1) $\Omega_2/2\pi$ = 1 MHz, (2) $\Omega_2/2\pi$ = 2 MHz, and (3) $\Omega_2/2\pi$ = 3 MHz. Dashed lines indicate the positions of the carrier (black), motion removing (red), and motion adding (blue) sideband transitions. Parameters: $\eta = 0.01$, $\omega_0/2\pi = 1.1$ MHz, $\Omega_1/2\pi = 17$ MHz, $\Delta_d/2\pi = 70$ MHz, and $\Gamma/2\pi = 20$ MHz with $\Gamma_1/\Gamma_2 = 1$. }
\label{eit_abs}
\end{figure}

At first glance, it may seem intuitive to estimate the cooling rate based on the absorption profile of the rest ion, where we calculate the steady state population of the excited state while varying the value of the probe detuning $\Delta_2$, as shown in Fig.\ref{eit_abs}(a). In particular, a higher red-to-blue sideband absorption ratio, red (blue) corresponding to motion-removing (adding) sideband, should suggest faster cooling. However, this holds true only when one laser is significantly weaker than the other ($\Omega_2 \ll \Omega_1$) \cite{RevModPhys.75.281}. Under such conditions, the cooling rate can be approximated by
\begin{equation}\label{wabs}
    W_{\text{abs}} \approx \eta^2 \Gamma \left(\rho_{ee,r}^{(0)} - \rho^{(0)}_{ee,b}\right),
\end{equation}
where $\Gamma = \Gamma_1 + \Gamma_2$ and $\rho^{(0)}_{ee,r}$ ($\rho^{(0)}_{ee,b}$) is the excited state population of the rest ion at the red (blue) sideband.

Plotting the absorption profile in Fig.\ref{eit_abs}(a) for three sets of parameters with $\Omega_2 < \Omega_1$ and the corresponding cooling curves in Fig.\ref{eit_abs}(b), we observe that the red-to-blue sideband absorption ratio qualitatively estimates the corresponding cooling rates, namely a larger ratio predicts faster cooling. However, the cooling rate prediction from the absorption profile is unreliable for arbitrary $\Omega_2/\Omega_1$ ratios, as we demonstrate next.

\subsubsection{Phonon rate equation}

When the internal (electronic) dynamics reach a steady state in a timescale much faster than the cooling dynamics of the motion, the former can be adiabatically eliminated to obtain a rate equation for the mean phonon number $\langle n \rangle$~\cite{Morigi_eit,Cirac_Zoller_cooling,RevModPhys.58.699}. This effective rate equation is given by 
\begin{equation}
  \langle \dot{n} \rangle = -W\langle n \rangle + \eta^2 A_{+}, 
\end{equation}
where   
\begin{flalign}\label{cooling_gio}
    W &= A_{-}-A_{+}, \nonumber \\
    A_{\pm} &= 2\text{Re}[S(\mp \omega_0)]. 
\end{flalign}
Here, $W$ is the cooling rate and  $A_{\pm}$ characterize the sideband excitation strengths. The quantity  $S(\text{v})$ is the value of the fluctuation spectrum \cite{Morigi_eit,Cirac_Zoller_cooling} of the operator $V_1$ at frequency v and is given by, 
\begin{flalign}\label{fluctuation spectrum}
    S(\text{v}) &= \text{Tr}_{\text{int}}\{V_1\left(-\mathcal{L}_0 -i\text{v}\right)^{-1}V_1 \rho_{\text{st}})\},
\end{flalign}
where the trace is taken over the internal degrees of freedom. $\mathcal{L}_0$ is the Lindblad superoperator of the rest ion under the dark state condition $\Delta_d$,  i.e.,
\begin{equation}
   \mathcal{L}_0 \rho = -i[H_R,\rho] + \mathcal{L}\rho,
\end{equation}  
and $\rho_{\text{st}}$ is the corresponding steady-state density matrix for an ion at rest. 

The system reaches the minimum steady-state value of $\langle n \rangle$ when $\Omega_1^2 + \Omega_2^2 = 4\omega_0(\omega_0 + \Delta_d)$. Under this condition, the maximum cooling rate scales as \cite{Morigi_eit}
\begin{equation}
    W_{\text{max}} \sim \eta^2 \left(\Omega_1 \Omega_2/\sqrt{\Omega_1^2 + \Omega_2^2}\right)^2/\Gamma.
\end{equation}
Thus, cooling is optimal at $\Omega_1 = \Omega_2$. 

\begin{figure}[htpb]
\includegraphics[scale=1.0]{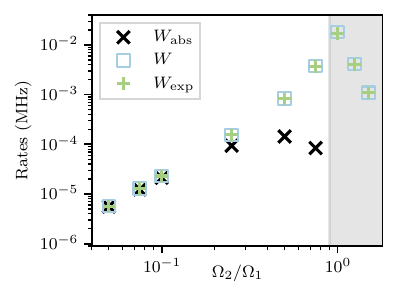}
\centering
\caption{Single EIT cooling: A comparison of the performance of the cooling rate estimation methods for varying Rabi frequency ratios $\Omega_2/\Omega_1$. $W_{abs}$ is the rest ion rate, Eq.\eqref{wabs}, $W$ denotes the result from the mean phonon rate equation, Eq.\eqref{cooling_gio}, and
$W_{\text{exp}}$ corresponds to the fitted, Eq.\eqref{exp_decay}, decay rate to numerical cooling dynamics, which is used to benchmark the methods. The shaded region indicates where $W_{\text{abs}}$ takes negative values. Note that $W_{\text{abs}}$ has been multiplied by a factor of 2 to match the other methods \cite{factor_2}. Parameters: $\eta = 0.01$, $\omega_0/2\pi = 2$ MHz, $\Omega_1/2\pi = 17$ MHz, $\Delta_d/2\pi = 70$ MHz, and $\Gamma/2\pi = 20$ MHz with $\Gamma_1/\Gamma_2 = 1$.}
\label{Rabi_ratio}
\end{figure}

In Fig.~\ref{Rabi_ratio}, we compare $W_{abs}$ and $W$ for various values of $\Omega_2/\Omega_1$ with the cooling rate $W_{exp}$ computed from the full dynamics, which we use to represent the correct dynamics. We find that $W_{abs}$ deviates significantly from the correct rate as $\Omega_2/\Omega_1$ increases. Furthermore, it fails to predict cooling near the optimal $\Omega_2/\Omega_1=1$, as $W_{abs}$ becomes negative. In contrast, $W$ remains in agreement with $W_{exp}$ for all $\Omega_2/\Omega_1$ values.

\subsubsection{Fictitious lasers}

\begin{figure*}[t!] 
\includegraphics[scale=1.]{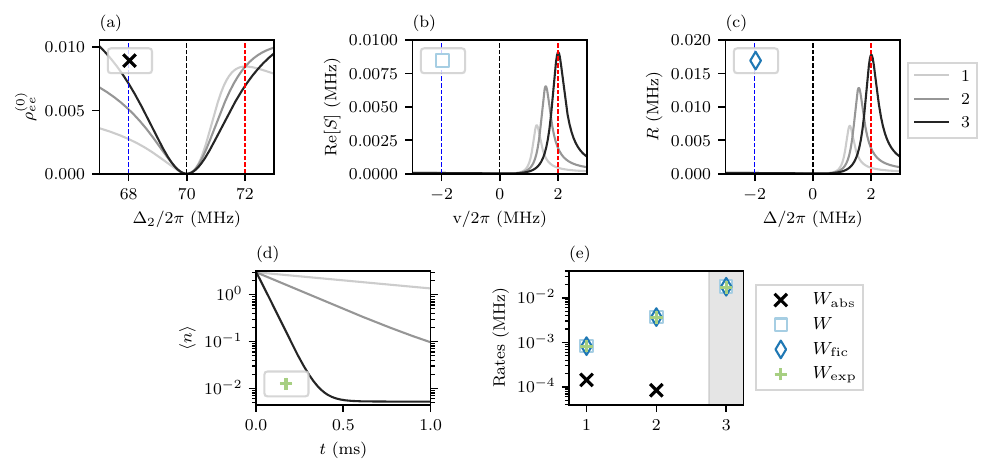}
\centering
\caption{Absorption profiles for single EIT cooling, based on (a) the steady-state population of the rest ion's excited state, (b) the real part of the fluctuation spectrum of $V_1$, (c) the absorption rate of fictitious lasers. Dashed lines indicate the positions of the carrier (black), motion removing (red), and motion adding (blue) sideband transitions. (d) The corresponding cooling dynamics for three scenarios:  (1) $\Omega_2/\Omega_1 = 0.5$, (2) $\Omega_2/\Omega_1 = 0.75$, and (3) $\Omega_2/\Omega_1 = 1$. (e) Cooling rates across all cases, where $W_{\text{abs}}$, $W$, $W_{\text{fic}}$ are obtained by spectra (a),(b), (c), respectively, as described in the main text. $W_{\text{exp}}$ is obtained by fiitng Eq.\eqref{exp_decay} to the cooling dynamics. The shaded region highlights negative values of $W_{\text{abs}}$. Note that $W_{\text{abs}}$ has been multiplied by a factor 2. Parameters: $\eta = 0.01$, $\omega_0/2\pi = 2$ MHz, $\Omega_1/2\pi = 17$ MHz, $\Delta_d/2\pi = 70$ MHz, nd $\Gamma/2\pi = 20$ MHz with $\Gamma_1/\Gamma_2 = 1$.}
\label{abs_all}
\end{figure*}

While $W$ accurately predicts the cooling rate even where $W_{abs}$ fails, its validity is restricted to the so-called weak sideband coupling (WSC) regime where the timescale for internal relaxation is much shorter than that for motional cooling. Inspired by a brief discussion in Ref.~\cite{Morigi_eit}, we consider a third approach to estimate the cooling rate that is intuitive and whose result agrees with $W$ in the WSC regime. Furthermore, it also provides qualitatively correct cooling rate estimates well beyond the WSC regime. 

Our approach is to treat the sideband transitions appearing in Eq.~\eqref{ld_eq} as additional `fictitious' lasers at frequencies $\widetilde{\omega}_j$ and Rabi frequencies $i\eta_j\Omega_j$ that probe the dressed states formed by the rest ion and the drive lasers. The cooling rate is then related to the absorption rate from the fictitious probe lasers when their frequencies are detuned $\pm \omega_0$ from the drive lasers.   

To formalize this intuition, we drop the motional degrees of freedom and write the ion's Hamiltonian in the rotating frame of the drive laser frequencies $\omega_j$ as, 
\begin{equation}
    H_F = H_R  + H_{-1} e^{-i\Delta t} + H_1e^{i\Delta t},
\end{equation}
with
\begin{flalign}
   H_{-1} &= \frac{i\eta\Omega_1}{2}\sigma_{e,g_{1}} -\frac{i\eta\Omega_2}{2}\sigma_{e,g_{2}}, \\ \nonumber
     H_1 &= H_{-1}^{\dagger}
\end{flalign}
where $H_R$ is defined at the dark condition $\Delta_d$ according to Eq.~\ref{restion_Ham}, the fictitious probe laser frequencies $\widetilde{\omega}_j$ are chosen such that the detunings $\Delta = \widetilde{\omega}_j - \omega_j$ are identical for both $\ket{g_j} \rightarrow \ket{e}$ transitions. In particular, $\Delta = +\omega_0$ ($\Delta = -\omega_0$) corresponds to the motion removing (adding) sideband. The ion's internal dynamics are then described by the master equation
\begin{equation}\label{fokker_}
    \dot{\rho} = ( \mathcal{L}_0 + \mathcal{L}_{-1}e^{-i\Delta t}  + \mathcal{L}_1e^{i\Delta t} ) \rho,
\end{equation}
with
\begin{flalign}
    \mathcal{L}_{-1} \rho & = -i[H_{-1},\rho], \nonumber \\
    \mathcal{L}_1 \rho &= -i[H_1,\rho].
\end{flalign}

We use the ansatz proposed in Ref.~\cite{Steck} to solve Eq.~\eqref{fokker_}, viz.,
\begin{equation}\label{sol_fic}
    \rho(t) \approx \rho^{(0)} + \delta \rho^{(0)} + \delta \rho^{(+)}e^{-i\Delta t} + \delta \rho^{(-)}e^{i \Delta t},
\end{equation}
where $\rho^{(0)}$ is the steady-state solution governed by $\mathcal{L}_0$, $\delta \rho^{(0)}$ is the correction due to fictitious lasers, and $\delta \rho^{(\pm)}$ represent the oscillatory corrections over time. It can be shown (see App.\ref{App_A}) that the steady-state photon absorption rate of the fictitious lasers is
\begin{flalign}\label{abs_rate}
    R(\Delta) = \, &\frac{\eta\Omega_1}{2}\left(\delta\rho_{eg_{1}}^{(+)} + \delta\rho_{g_{1}e}^{(-)}\right) \nonumber \\
    - &\frac{\eta\Omega_2}{2}\left(\delta\rho_{eg_{2}}^{(+)} + \delta\rho_{g_{2}e}^{(-)}\right),
\end{flalign}
where $\delta \rho_{\mu \nu}^{(\pm)}$ are matrix elements of $\delta \rho^{(\pm)}$. Then, a cooling rate can be determined from the absorption rate of the fictitious lasers at the motional sideband frequencies as
\begin{equation}\label{fic_rate}
  W_{\text{fic}} = R(\omega_0) - R(-\omega_0).  
\end{equation}

In Fig.~\ref{abs_all}, we study the ability of all three methods to accurately estimate the cooling rate. Figures~\ref{abs_all}(a), (b) and (c) show respectively the absorption profile of the rest ion, the fluctuation spectrum of $V_1$ and the absorption rate from the fictitious lasers for three parameter sets around and at the optimal $\Omega_2/\Omega_1$ ratio. The corresponding cooling curves obtained from full dynamics simulations are plotted in Fig.\ref{abs_all}(d), allowing us to extract the numerical cooling rate $W_{exp}$ based on which we make our assessment.  

If the absorption profile in Fig.\ref{abs_all}(a) were a reliable indicator, the first case (parameter set) would result in the most rapid cooling, followed by the second, while the third case would result in heating based on the red-to-blue sideband absorption ratio. However, the numerical results contradict these expectations. On the other hand, the fluctuation spectrum of $V_1$ and the absorption rate from the fictitious lasers not only correctly predict the qualitative cooling behaviour according to the corresponding red-to-blue sideband absorption ratios, but also provide quantitatively precise estimates, as demonstrated in Fig.\ref{abs_all}(e).

\vspace{3mm}
All optimization approaches discussed thus far concern systems operating within the LD regime. We note that in the context of EIT cooling, including multilevel schemes, efforts have recently been directed toward developing semiclassical methods to predict cooling rates beyond the LD regime \cite{bartolotta2024lasercoolingtrappedioncrystal}.

\subsection{Sideband coupling regimes}

So far, our analysis has focused on cooling in the WSC regime, $\eta\Omega \ll \Gamma$, where the internal dynamics occur on a faster timescale than the motional cooling \cite{Morigi_eit,Zhang_2021}. When $\eta\Omega \gtrsim \Gamma$, the system is in a strong sideband coupling (SSC) regime and the timescales for internal and motional dynamics start to become comparable. Making accurate predictions of cooling rates is challenging in this regime since the internal dynamics can no longer be adiabatically eliminated.

\begin{figure}[htpb]
\includegraphics[width=\linewidth]{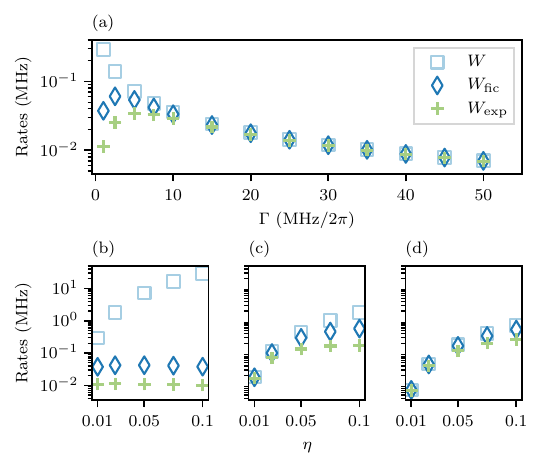}
\centering
\caption{Single EIT cooling: (a) Cooling rates for different decay rates at $\eta = 0.01$. Second row: Cooling rates as functions of the Lamb-Dicke parameter $\eta$ for (b) $\Gamma/2\pi = 1$ MHz, (c) $\Gamma/2\pi = 20$ MHz, and (d) $\Gamma/2\pi = 50$ MHz. $W$ represents the rate equation result, Eq.\eqref{cooling_gio}, $W_{\text{fic}}$ corresponds to the fictitious laser absorption rate, Eq.\eqref{fic_rate}, and $W_{\text{exp}}$ is the numerically calculated rate. Parameters: $\omega_0/2\pi = 2$ MHz, $\Omega_1/2\pi = \Omega_2/2\pi = 17$ MHz, $\Delta_d/2\pi = 70$ MHz, and $\Gamma_1/\Gamma_2 = 1$.}
\label{adia_val}
\end{figure}

In Fig.~\ref{adia_val}, we explore the crossover from the WSC to SSC. In Fig.\ref{adia_val}(a) we hold $\eta\Omega$ constant while sweeping the decay rate $\Gamma$. When the coupling strength $\eta \Omega$ is comparable to $\Gamma$, both $W$ and $W_{fic}$ diverge from the numerically calculated cooling rate $W_{exp}$, marking the breakdown of the validity of adiabatic elimination, as the system is in the SSC regime. Once $\eta \Omega$ becomes smaller than $\Gamma$ the system transitions back to the WSC regime where both $W$ and $W_{fic}$ agree with $W_{exp}$.

In Fig.~\ref{adia_val} (b),(c) and (d) we tune the coupling strength by varying the LD parameter $\eta$ for three constant values of decay rate $\Gamma/2\pi = 1$ MHz, $\Gamma/2\pi = 20$ MHz, and $\Gamma/2\pi = 50$ MHz, respectively. For the latter two cases, the system is in the WSC regime where $W$ and $W_{fic}$ are in reasonable agreement and also approximately match $W_{exp}$. The $\eta^2$ dependence of the cooling rate is also evident in these cases. However, for $\Gamma/2\pi = 1$ MHz, the system is in the SSC regime and the full dynamics reveals that $W_{exp}$ is nearly independent of $\eta$. This observation is in agreement with the analytical cooling rates reported for EIT cooling in the SSC regime in Ref.~\cite{Zhang_2021}. In contrast, here the rate $W$ is derived under a WSC approximation and always predicts an $\eta^2$ scaling, and therefore does not capture the cooling rates even qualitatively in the SSC regime. 

Notably, while neither $W$ nor $W_{fic}$ quantitatively predict cooling rates the SSC regime,  $W_{fic}$ yields qualitatively correct results. This can be understood by considering the role of the internal dynamics. In the WSC regime, rapid internal relaxation and weak atom-motion coupling ensure that the rest ion's steady state $\rho^{(0)}$ and its perturbative correction  $\rho^{(0)} + \delta\rho^{(0)}$ (Eq.~\ref{sol_fic}) accurately approximate the full system's internal dynamics during cooling. As we depart from the WSC regime, internal and cooling dynamics occur on similar or slower timescales, making $\rho^{(0)}$ unreliable approximation of internal dynamics. This explains the failure of $W$ in the SSC regime.

The success of $W_{fic}$ in yielding qualitatively agreement is attributed to the fact that $\rho^{(0)} + \delta\rho^{(0)}$ approximates the full system's average internal dynamics during cooling. This approximation remains valid as long as the atom-motion coupling is weak enough such that  the internal dynamics of the rest ion provide a reasonable approximation to the internal dynamics of the full system. By incorporating fictitious lasers into the rest ion dynamics, we effectively approximate the absorption at the sidebands as a perturbation. Since the absorption at the sidebands is negligible once the motional ground state is reached, the fictitious lasers approximate the population's dynamics for $\langle n \rangle \neq$ 0. This explains why $W_{fic}$ outperforms W in the SSC regime.

Note that as the sideband coupling increases, the system will eventually enter a regime in which the internal dynamics of the full system deviate significantly from those of the rest ion. Under these conditions, $\rho^{(0)} + \delta\rho^{(0)}$ no longer serves as a reliable approximation of the internal dynamics during cooling. While this regime is not observed for the minimal three level system and the parameter range examined in this section, such deviations will become apparent once we extend our analysis to multilevel systems.

\section{MULTILEVEL SYSTEM}

We now study the ability of the three methods discussed in Sec.~II to predict the cooling rates for ions progressively involving a larger number of electronic levels. In Appendix~\ref{double_eit}, we show a good quantitative agreement of both $W$ and $W_{fic}$ with the full dynamics results for the case of double EIT cooling~\cite{double_eit_theory1,double_eit_theory2,Semerikov2018,PhysRevLett.125.053001,Qiao2021}. Here, we turn to analyze the cooling of ions with non-zero nuclear spin, which leads to ground and excited state manifolds with several degenerate levels in each manifold. The degeneracy is typically lifted by application of a magnetic field $B$. To study a variety of trapped ion species with this generic level structure, we have developed a code that can take a variable nuclear spin $I$ and electronic angular momenta $J$ as inputs to construct the relevant electronic level structure, compute Clebsch-Gordan coefficients and set up the laser cooling master equation for a given laser configuration, which we subsequently solve using QuTiP \cite{qutip1,qutip2,qutip3}. Our code is publicly available \cite{github} and in Appendix~\ref{dyn_multi}, we summarize the atomic physics concepts involved in implementing it. 

As in single EIT cooling, the objective is to cancel the carier transition and to find the parameters that maximize the cooling rate without explicitly calculating the full dynamics. A common approach for cooling optimization involves varying the Rabi frequencies of the probe and drive lasers, one at a time, and examining their impact on the numerically calculated final mean phonon number \cite{Qiao2021,PhysRevLett.133.113204,wu2024}.
However, this method does not ensure a thorough exploration of the parameter space and can be computationally demanding, particularly in multilevel systems. Instead, in this section we evaluate the cooling estimation techniques developed for single EIT, as they can easily be extended for multilevel systems (App.\ref{cool_multi}).

\subsection{Model}
\begin{figure}[htpb!] 
\includegraphics[width=\linewidth]{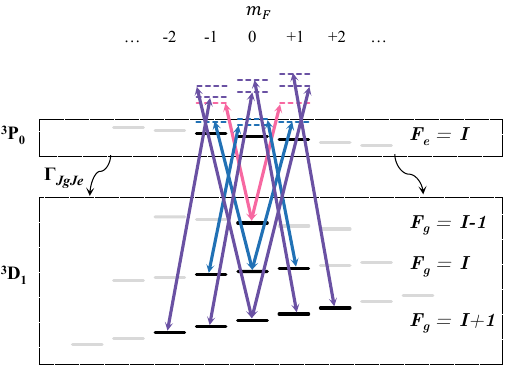}
\centering
\caption{Cooling transition $^{3}$D$_1$ $\leftrightarrow$ $^{3}$P$_0$ of $^{176}$Lu$^+$. Bold black lines (hyperfine states) and bold arrows (laser transitions) are sketched for a toy model with nuclear spin $I$ = 1. Each $F_g \rightarrow F_e$ transition is driven by a single laser frequency with both $\sigma^+$ and $\sigma^-$ polarization components. The colored dashed lines show the detuning for each individual $\ket{F_g, m_{F_g}} \rightarrow \ket{F_e, m_{F_e}}$ transition. Athlough decay between the excited and ground hyperfine states is included in the model, only the total spontaneous decay rate $\Gamma_{J_gJ_e}$ is illustrated here for simplicity. The grey lines indicate how the level structure scales with increasing $I$. Laser transitions between grey levels are omitted for clarity, but they are included in our model.}
\label{multilevel}
\end{figure}

As a reference for a multilevel cooling transition, we employ the $^{3}$D$_1$ $\leftrightarrow$ $^{3}$P$_0$ transition of $^{176}$Lu$^+$ \cite{Barrett_2015, PhysRevA.92.032108, PhysRevLett.117.160802, Arnold2018, PhysRevA.98.022509, PhysRevLett.124.083202, doi:10.1126/sciadv.adg1971} which consists of a dissipative excited state manifold and long-lived ground state manifolds. We construct toy models from this reference model by fixing $J_e=0$, $J_g=1$ and varying the nuclear spin $I$, see Fig.\ref{multilevel}. We address the system with three different laser frequencies: the ground state manifolds $F_g = I+1$ and $F_g = I-1$ are each coupled to the excited states $F_e$ by a single driving beam respectively, while the $F_g = I \rightarrow F_e$ transition is addressed by a probe beam, each with a Rabi frequency $\Omega_{F_gF_e}$ and detuning $\Delta_{F_gF_e}$ defined while ignoring the hyperfine structure. Rabi frequencies between hyperfine states $\ket{F_g, m_{F_g}} \rightarrow \ket{F_e, m_{F_e}}$ are appropriately calculated using the Wigner-Eckart theorem \cite{atomic} while their detunings are corrected from $\Delta_{F_gF_e}$ to account for the Zeeman shifts. All beams are assumed to have $\sigma_{+}$ and $\sigma_{-}$ polarization components. The laser driving the $F_g = I+1 \rightarrow F_e$ transition counterpropagates relative to the other two. Due to the model's complexilty, the master equation for laser cooling under such a configuration is described in Appendix~\ref{dyn_multi} rather than in the main text.

\subsubsection{Dark states}

In lambda and tripod level systems, dark states are formed when the two detunings become equal. However, equal detunings alone do not  always guarantee the existence of dark states \cite{Multilevel_dark}. A dark state satisfies 
\begin{equation}\label{dark_eq}
\mathcal{K}\rho=0,
\end{equation}
where $\dot{\rho}=\mathcal{K}\rho$, with no population in the excited state manifold. It can be shown \cite{Multilevel_dark} that Eq.\eqref{dark_eq} is equivalent to
\begin{flalign}
    &[PHP,\rho] = 0, \label{detu_con} \\
    &QHP\rho = 0, \label{rabi_con}
\end{flalign}
where $P$ and $Q$ are the projection operators spanning the ground states and excited states of the system, respectively. Eq.\eqref{detu_con} has non-trivial solutions if the spectrum of $PHP$ is degenerate. We note that the Hamiltonian is assumed to be time-independent. For the Hamiltonian in Eq.\eqref{Ham_multi_inter} decribing our model, this degeneracy correspond to enforcing equal detunings between $\ket{F_{g}\, m_{F_g}} \rightarrow \ket{F_e\, m_{F_e}}$ transitions. However, this condition alone is not sufficient as Eq.\eqref{rabi_con} must also be satisfied.

If there are $d_s > 1$ equal detunings associated with ground states $\ket{\Phi_g\, m_{\Phi_g}}$, we define $P_s$ as the orthogonal projector spanning these $d_s$ states. Then Eq.\eqref{rabi_con} will be satisfied provided that $\text{dim}[\text{ker}[QHP_s]]\geq 1$ which can be equivalently expressed as
\begin{equation}\label{dark_cond}
    \text{rank}[QHP_s] \leq d_s - 1.
\end{equation}
Therefore, as long as Eq.\eqref{dark_cond}
holds, equal detunings are suffcient to establish dark states. This is true for both lambda and tripod systems. Nevertheless, when Eq.\eqref{dark_cond} is no longer met, we need to impose conditions on the Rabi frequencies such that $\text{dim}[\text{ker}[QHP_s]] \neq 0$.

In our system we can identify dark states  by calculating the absorption profile of the rest ion while scanning the detuning of the probe as illustrated in Fig.\ref{abs_I1_all}(a) and Fig.\ref{abs_I7_all}(a). For $\sigma^{+}$ and $\sigma^{-}$ laser polarization components, we find dark states that are independent of the Rabi frequencies. For example, for I=1 when $\Delta_{11}/2\pi$ = 20.38 MHz or 20.5 MHz,  $\text{rank}[QHP_s]$ = 1 and $d_s$=2, so Eq.\eqref{dark_cond} is met . However, we note that in experimental set ups where there are misaligments between the laser propagation and the magnetic field some $\pi$ polarization might be introduced and it will affect our cooling as we describe in App.\ref{App:polarization}.

\subsection{I=1}

\begin{figure*}[htbp!] 
\includegraphics[scale=1.0]{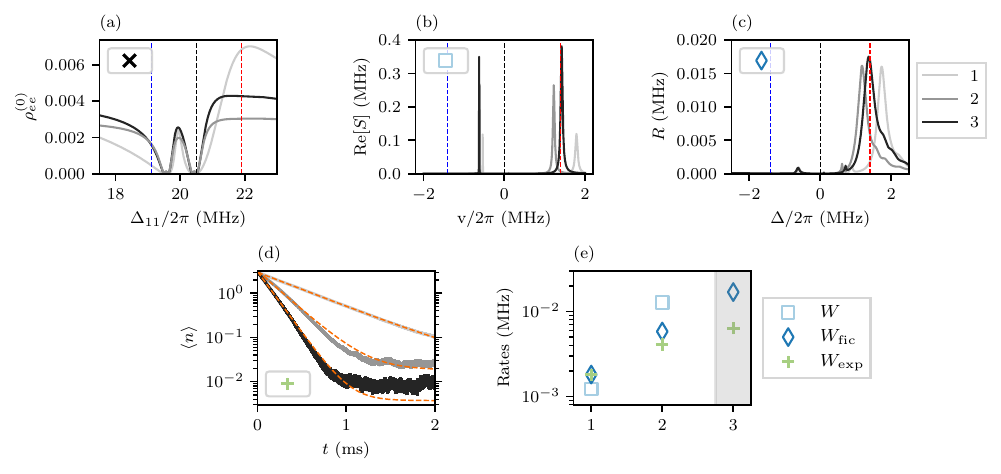}
\centering
\caption{Absorption profiles for multilevel EIT cooling, $I$ = 1 based on (a) the sum of the steady-state population of the rest ion's excited states, (b) the real part of the fluctuation spectrum of $\widetilde{V}_1$, (c) the absorption rate of fictitious lasers for three cases. As before, dashed lines indicate the positions of the carrier (black), motion removing (red), and motion adding (blue) sideband transitions. (d) The corresponding cooling dynamics averaged over 500 quantum trajectories where the line width encodes the standard error. Orange dashed line represents fitted exponential decay Eq.\ref{exp_decay} from which we obtain $W_{\text{exp}}$. (e) Cooling rates across all cases, where $W$, $W_{\text{fic}}$ are obtained from spectra (b) and (c), respectively, as described in the main text.  $W_{\text{abs}}$ obtained from absorption profile (a) is of the order of $10^{-5}-10^{-4}$ MHz for all cases and is not depicted here. In the shaded area $W$ assumes a value of $\approx$ 0.4 MHz and is dropped for clarity. Parameters used: (1) $\Omega_{01}/2\pi = \Omega_{21}/2\pi$ = 30 MHz, $\Omega_{11}/2\pi$ = 5 MHz (2) $\Omega_{01}/2\pi = \Omega_{21}/2\pi$ = 23 MHz, $\Omega_{11}/2\pi$ = 8 MHz (3) $\Omega_{01}/2\pi = \Omega_{21}/2\pi$ = 23.5 MHz, $\Omega_{11}/2\pi$ = 10 MHz ($\Omega_{F_g F_e}$). For all: $\eta = 0.044$, $\omega_0/2\pi = 1.4 $ MHz, B = 5 G, $\Delta_{01}/2\pi = 10$ MHz, $\Delta_{11}/2\pi = 20.5$ MHz, $\Delta_{21}/2\pi = 20$ MHz ($\Delta_{F_g F_e}$) and $\Gamma_{J_gJ_e}/2\pi = 2.45 $ MHz.}
\label{abs_I1_all}
\end{figure*}
We begin by examining a toy model with nuclear spin $I=1$. This model is obtained by truncating the full $\rm ^{176}Lu^+$ level structure, as shown in Fig.~\ref{multilevel}. Following the methodology we adopted for single EIT cooling, we calculate the absorption profile of the rest ion in Fig.\ref{abs_I1_all}(a) for three distinct sets of laser parameters, summing over all excited state populations. The dark steady-states can be easily identified as the points where the total excited state population vanishes. They are characterized by the value of the probe laser $\Delta_{11}$. The fluctuation spectrum of $\widetilde{V}_1$, Fig.\ref{abs_I1_all}(b) and the absorption rate of the fictitious lasers driving the sidebands, Fig.\ref{abs_I1_all}(c), are also calculated to estimate the cooling rates. It is important to note that the chosen dark state is unique, given that the cooling optimization methods depend on steady state calculations.

For the full dynamics simulations, the motion of the ion is assumed to be initialized in a thermal state with mean phonon number $\langle n \rangle$ = 3 by a prior Doppler cooling cycle and the atomic state is assumed to be initialized in $\ket{F_g = 2, m_{Fg} = -2}$ by optical pumping. The cooling dynamics [Eq.\eqref{dynamics_multi}] are simulated by averaging over 500 quantum trajectories \cite{quantum_trajectories,fancy_qt}. The cooling curves for the various parameter sets are shown in Fig. \ref{abs_I1_all}(d).

We can now assess the ability of the three estimation techniques in qualitatively and quantitatively predicting the cooling rates. A higher asymmetry in the red-to-blue sideband heights indicates a faster cooling rate. 
The absorption profile of the rest ion in Fig.\ref{abs_I1_all}(a)
incorrectly predicts that the rates follow the order case $1$ $>$ case $3$ $>$ case $2$, while the correct order from the full dynamics calculations is case $3$ $>$ case $2$ $>$ case $1$. In contrast, both the fluctuation spectrum [Fig.\ref{abs_I1_all}(b)] and the absorption rate from the fictitious lasers [Fig.\ref{abs_I1_all}(c)] get the order correct, and hence are in qualitative agreement with the full dynamics results. For a more quantitative comparison, Fig.\ref{abs_I1_all}(e) compares the cooling rates $W$ and $W_{fic}$ with each other and with $W_{exp}$. The former two do not coincide, signaling that the chosen parameter sets are not in the WSC regime. Moreover, for cases $2$ and $3$, neither $W$ nor $W_{fic}$ coincide with $W_{exp}$. These results are along the same lines as our observations in the case of single EIT cooling outside the WSC regime. 

\begin{figure*}[htbp!]
\includegraphics[scale=1.0]{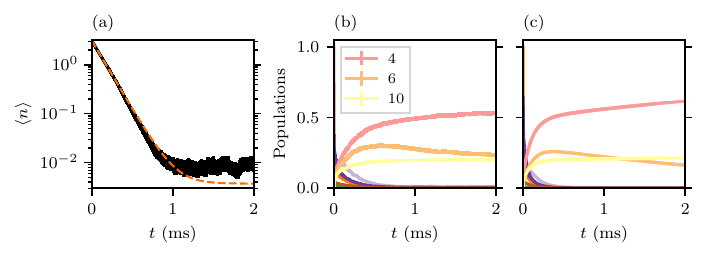}
\centering
\caption{$I$ = 1: (a) Cooling dynamics where orange dashed line represents the fitted exponential decay Eq.\ref{exp_decay}, (b) respective internal dynamics of the full system averaged over 500 quantum trajectories for $\eta = 0.044$ and (c) internal dynamics of the rest ion $\eta$ = 0. Parameters: $\omega_0/2\pi = 1.4 $ MHz, B = 5 G, $\Omega_{01}/2\pi = \Omega_{21}/2\pi$ = 23.5 MHz, $\Omega_{11}/2\pi$ = 10 MHz ($\Omega_{F_g F_e}$), $\Delta_{01}/2\pi = 10$ MHz, $\Delta_{11}/2\pi = 20.5$ MHz, $\Delta_{21}/2\pi = 20$ MHz ($\Delta_{F_g F_e}$) and $\Gamma_{J_gJ_e}/2\pi = 2.45 $ MHz. In (a) and (b) the lines' width denotes the trajectories' standard error. States are enumerated by increasing $m_F$ number beginning with 0: $\ket{F_e, m_{Fe}=-F_e}$, then 4: $\ket{F_g=1,m_{Fg}=-1}$, 6: $\ket{F_g=1,m_{Fg}=1}$, 10: $\ket{F_g=2,m_{Fg}=1}$. Full system's steady-state populations: 4 $\approx$ 0.74, 6 $\approx$ 0.04, 10 $\approx$ 0.22. Rest ion's steady-state populations: 4 $\approx$ 0.77, 6 $\approx$ 0, 10 $\approx$ 0.23.}
\label{I1_pop}
\end{figure*}

Nevertheless, $W_{fic}$ is in the same ballpark as $W_{exp}$ for all cases, unlike $W$, despite the system operating outside the WSC regime for the selected parameters. In Fig.~\ref{I1_pop}, we provide some insight into why this is the case by plotting (a) the cooling curve, (b) the internal dynamics for the full system  and (c) the internal dynamics for the rest ion for case $3$. We observe that the motional ground state is prepared well before the atomic populations reach their steady state, which is expected for a system operating in a SSC regime.

Although the steady state populations of the full system and the rest ion exhibit close agreement, with an Euclidean distance of 0.05, these populations are insufficient approximations of the average populations during cooling. Specifically, comparing the averaged populations of the full system (over 0.5 ms of cooling) with the rest ion's steady state populations, we find an increased distance of 0.55, which explains the deviations of $W$ from the numerical results.

 On the other hand,  the fictitious lasers method's steady state $\rho^{(0)} + \delta\rho^{(0)}$ provides a more accurate approximation of the average populations. This is attributed to the similarity between the internal dynamics of the full system and those of the rest ion. The distance between the populations obtained from $\rho^{(0)} + \delta\rho^{(0)}$ and the steady state of the internal dynamics of the full system is 0.74. However, the distance from the average populations is 0.34. This improvement likely explains why $W_{fic}$ yields qualitatively better results than $W$.

\subsubsection{Sideband coupling regimes}

\begin{figure}[htbp!]
\includegraphics[width=\linewidth]{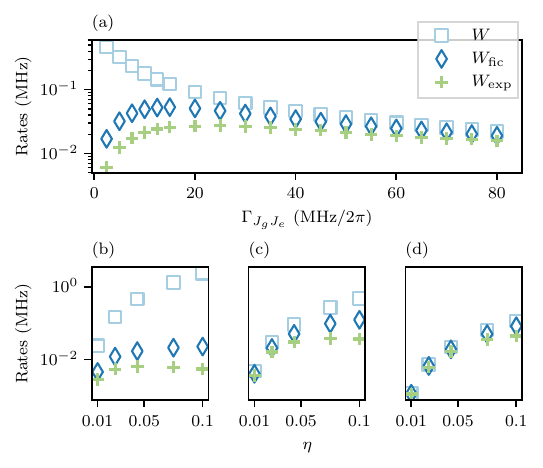}
\centering
\caption{$I$ = 1: (a) Cooling rates for different decay rates at $\eta = 0.044$. Second row: Cooling rates as functions of the Lamb-Dicke parameter $\eta$ for (b) $\Gamma_{J_gJ_e}/2\pi = 2.45$ MHz, (c) $\Gamma_{J_gJ_e}/2\pi = 20$ MHz, and (d) $\Gamma_{J_gJ_e}/2\pi = 80$ MHz. $W$ represents the rate equation result, $W_{\text{fic}}$ corresponds to the fictitious laser absorption rate, and $W_{\text{exp}}$ is the numerically calculated result. Parameters: $\omega_0/2\pi = 1.4$ MHz, $B = $ 5 G, $\Omega_{01}/2\pi = \Omega_{21}/2\pi$ = 23.5 MHz, $\Omega_{11}/2\pi$ = 10 MHz ($\Omega_{F_g F_e}$), $\Delta_{01}/2\pi = 10$ MHz, $\Delta_{11}/2\pi = 20.5$ MHz, $\Delta_{21}/2\pi = 20$ MHz ($\Delta_{F_g F_e}$).}
\label{adia_val_I1}
\end{figure}

We next explore the transition from weak to strong sideband coupling regimes. In Fig.\ref{adia_val_I1}(a), we keep $\eta\Omega_{F_gF_e}$ constant while varying the decay rate $\Gamma_{J_gJ_e}$. For small values of $\Gamma_{J_gJ_e}$, the cooling rates $W$ and $W_{fic}$ deviate from each other and from the full dynamics result $W_{exp}$. However, as $\Gamma_{J_gJ_e}$ increases, the rates converge as the system transitions from the SSC to the WSC regime. We have verified that both the fluctuation spectrum and the absorption profile of the fictitious lasers are in good agreement with each other in the WSC regime, while the absorption profile of the rest ion deviates from the former two since the probe and drive lasers are not strongly imbalanced.

In Fig.\ref{adia_val_I1}(b),(c),(d) we compare $W$ and $W_{fic}$ against $W_{exp}$ for varying $\eta$ at fixed decay rates $\Gamma_{J_gJ_e}/2\pi = 2.45$ MHz,  $\Gamma_{J_gJ_e}/2\pi = 20$ MHz, and $\Gamma_{J_gJ_e}/2\pi = 80$ MHz, respectively. As in single EIT, we observe a disagreement from the $\eta^2$ dependency in the SSC regime and its recovery in the WSC regime. Although both $W$ and $W_{fic}$ deviate from the numerical rates $W_{exp}$ in the SSC regime, $W_{fic}$ remains a useful tool to make qualitative estimations, given that the internal dynamics of the full system resemble the rest ion ones. This allows $\rho^{(0)} + \delta\rho^{(0)}$ to make an effective approximation of the average populations during cooling.

\subsection{I=7}

\begin{figure*}[htbp!] 
\includegraphics[scale=1.0]{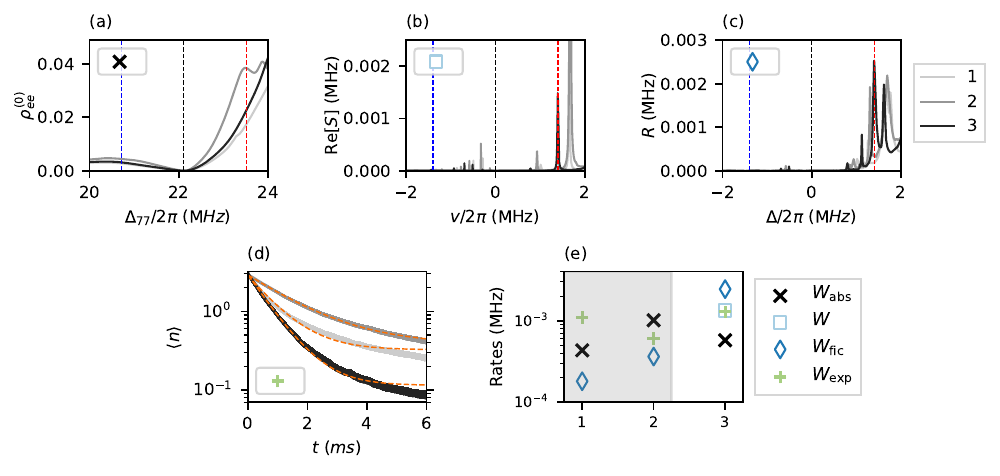}
\centering
\caption{Absorption profiles for multilevel EIT cooling, $I$ = 7 based on (a) the sum of the steady-state population of the rest ion's excited states, (b) the real part of the fluctuation spectrum of $\widetilde{V}_1$,  (c) the absorption rate of fictitious lasers for three scenarios. Dashed lines indicate the positions of the carrier (black), motion removing (red), and motion adding (blue) sideband transitions. (d) The corresponding cooling dynamics averaged over 500 quantum trajectories where the line width encodes the standard error. Orange dashed line represents fitted exponential decay Eq.\ref{exp_decay} from which we obtain $W_{\text{exp}}$. (e) Cooling rates across these cases, where $W_{\text{abs}}$, $W$, $W_{\text{fic}}$ are obtained from spectra (a), (b), (c), respectively. In the shaded area $W$ assumes values of the order of $10^{-6}$. Parameters used: (1) $\Omega_{67}/2\pi = \Omega_{87}/2\pi$ = 73 MHz, $\Omega_{77}/2\pi$ = 19 MHz (2) $\Omega_{67}/2\pi = \Omega_{87}/2\pi$ = 63 MHz, $\Omega_{77}/2\pi$ = 20 MHz (3) $\Omega_{67}/2\pi = \Omega_{87}/2\pi$ = 70 MHz, $\Omega_{77}/2\pi$ = 19 MHz ($\Omega_{F_g F_e}$). For all: $\eta = 0.044$, $\omega_0/2\pi = 1.4 $ MHz, B = 4 G, $\Delta_{67}/2\pi = 10$ MHz, $\Delta_{77}/2\pi = 22.1$ MHz, $\Delta_{87}/2\pi = 20$ MHz ($\Delta_{F_g F_e}$) and $\Gamma_{J_gJ_e}/2\pi = 2.45 $ MHz.}
\label{abs_I7_all}
\end{figure*}

We now extend our study to the full $^{176}$Lu$^+$ ($I=7$) level structure following the same approach. Dark states are identified based on the absorption profile of the rest ion, which is shown in Fig.\ref{abs_I7_all}(a) for three distinct sets of laser parameters. Choosing a unique dark state characterized by the value of the probe laser $\Delta_{77}$,  we calculate the fluctuation spectrum in Fig.\ref{abs_I7_all}(b) and the absorption rate of the fictitious lasers in Fig.\ref{abs_I7_all}(c). For the full dynamics simulations [Eq.\eqref{dynamics_multi}], the ion motion is  initialized in a thermal state characterized by $\langle n \rangle$ = 3 and the  atomic state is initialized in $\ket{F_g = 8, m_{Fg} = -8}$. The resulting cooling curves are shown in Fig.\ref{abs_I7_all}(d).

Comparing $W_{abs}$, $W$ and $W_{fic}$ with $W_{exp}$, we find that all cooling rate estimation techniques considered in this work are unreliable for this system for the chosen parameter sets. Achieving a larger red-to-blue sideband absorption ratio does not necessarily guarantee faster cooling. In Fig.\ref{abs_I7_all}, while the second case is expected to outperform the first one, based on the cooling rates $W_{fic}$ in Fig.\ref{abs_I7_all}(e), the numerical calculations contradict this expectation. The explicit cooling rates calculated in Fig.\ref{abs_I7_all}(e) verify operation outside of the WSC regime, as reflected by the disagreement between $W$ and $W_{fic}$. Moreover, the Fig.\ref{abs_I7_all}(e)  confirms that none of the estimation methods is able to correctly capture the qualitative trend of $W_{exp}$. We point out that the exponential decay fitting Eq.\eqref{exp_decay} in Fig.\ref{abs_I7_all}(e) is not optimal. However, it sufficiently captures the initial decay required to obtain $W_{exp}$. Slight deviations from an ideal exponential decay can be attributed to the internal dynamics.

\begin{figure*}[htbp!]
\includegraphics[scale=1.0]{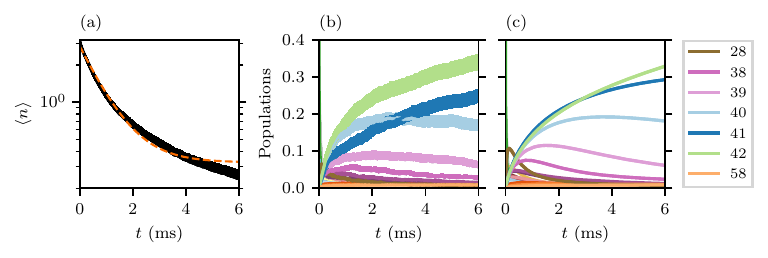}
\centering
\caption{$I$ = 7: (a) Cooling dynamics where the orange dashed line represents fitted exponential decay Eq.\ref{exp_decay} and (b) respective internal dynamics of the full system averaged over 500 quantum trajectories for $\eta = 0.044$ and (c) internal dynamics of the rest ion $\eta$ = 0. Parameters: $\omega_0/2\pi = 1.4 $ MHz, B = 4 G, $\Omega_{67}/2\pi = \Omega_{87}/2\pi$ = 73 MHz, $\Omega_{77}/2\pi$ = 19 MHz ($\Omega_{F_g F_e}$), $\Delta_{67}/2\pi = 10$ MHz, $\Delta_{77}/2\pi = 22.1$ MHz, $\Delta_{87}/2\pi = 20$ MHz ($\Delta_{F_g F_e}$) and $\Gamma_{J_gJ_e}/2\pi = 2.45 $ MHz. In (a) and (b) the lines' width denotes the trajectories' standard error. States are enumerated by increasing $m_F$ number beginning with 0: $\ket{F_e, m_{Fe}=-F_e}$, then 28: $\ket{F_g=7,m_{Fg}=-7}$, 38: $\ket{F_g=7,m_{Fg}=3}$, 39: $\ket{F_g=7,m_{Fg}=4}$, 40: $\ket{F_g=7,m_{Fg}=5}$, 41: $\ket{F_g=7,m_{Fg}=6}$, 42: $\ket{F_g=7,m_{Fg}=7}$, 58: $\ket{F_g=8,m_{Fg}=7}$. The width of the lines represents the standard error of the trajectories. Full system's steady-state populations: 39 $\approx$ 0.02, 40 $\approx$ 0.02, 41 $\approx$ 0.12,  42 $\approx$ 0.81, 58 $\approx$ 0.01. Rest ion's steady-state: 42 $\approx$ 0.99, 58 $\approx$ 0.01.}
\label{I7_pop}
\end{figure*}

In this system, the internal dynamics evolve on a significantly slower timescale than the cooling process. Unlike the case with $I$=1,~ $\rho^{(0)} + \delta\rho^{(0)}$ no longer serves as an accurate approximation of the internal dynamics during cooling, marking an ultra-strong sideband coupling (U-SSC) regime. As Fig.\ref{I7_pop} reveals, the internal dynamics of the full system [Fig.\ref{I7_pop}(b)] are no longer adequately approximated by the internal dynamics of the rest ion [Fig.\ref{I7_pop}(b)]. Consequently, $\rho^{(0)} + \delta \rho^{(0)}$ fails to yield a valid approximation of the average populations during cooling. 

In particular, for case 1 [Fig.\ref{I7_pop}], the steady state populations of the rest ion exhibit a distance 0.22 from the steady state populations of the full dynamics and a distance of 0.78 from the average populations during cooling. On the other hand, the populations derived from $\rho^{(0)} + \delta \rho^{(0)}$ exhibit a distance 0.15 from the steady state populations of the full system and a 0.69 distance from the average populations.

Systems operating in the U-SSC regime require a more careful treatment to predict the cooling rates that go beyond the techniques under consideration in this work. Nevertheless, as the cooling curves in Fig.~\ref{abs_I7_all}(d) demonstrate, multilevel EIT cooling remains feasible even in this
regime.

\section{CONCLUSIONS AND OUTLOOK}

We numerically studied EIT cooling in ions with multilevel electronic structure. We relaxed the necessity for isolation of a three-level system and exploited the inherent dark states of the system to perform cooling. For systems in the weak sideband coupling (WSC) regime where the internal dynamics reach their steady state in a timescale much faster than the timescale of the cooling, we demonstrated that cooling rates can be accurately estimated using both well-established methods ($W$) as well as an alternative method based on the absorption rate of fictitious lasers driving electronic transitions at the sideband frequencies ($W_{fic}$). Furthermore, we clarified the limitations of the absorption profile of the rest ion ($W_{abs}$)  as an indicator for cooling rate estimation. $W_{abs}$ can only offer a valid cooling rate estimation for systems that operate with a strong imbalance between Rabi frequencies. However, optimal cooling is achieved when the Rabi frequencies are of similar order.

Moreover, we introduced a tool to identify the sideband coupling regimes without explicitly simulating the full dynamics. Disagreements between $W$ and $W_{fic}$ indicate operation outside the WSC regime. Realistic multilevel systems are likely to operate in a strong sideband coupling regime where the relatively slow internal dynamics make cooling rate predictions challenging. When internal dynamics of the rest ion approximate the internal dynamics of the full system, $W_{fic}$ offers an efficient initial estimate of the cooling rate. On the contrary, in the ultra-strong sideband coupling (U-SSC) regime, where the sideband coupling causes the internal dynamics of the rest ion to deviate significantly from those of the full system, $W_{fic}$ fails to make accurate predictions. For such cases, a more nuanced approach to estimate the cooling rate without explicit calculation of the full dynamics is required, which we leave for future work. Nevertheless, our findings confirm that multilevel EIT cooling is feasible even when the internal dynamics evolve on much longer times than the cooling timescale. Our results are relevant for the current generation of trapped ion experiments, where a variety of exotic ion species with rich electronic structures are being explored for various quantum information applications.

\section*{ACKNOWLEDGEMENTS}
This work used the Dutch national e-infrastructure with the support of the
SURF Cooperative and the Dutch Research Council (NWO) using grant no.~EINF-8630. A.S.N. is supported by the Dutch Research Council
(NWO/OCW) as a part of the Quantum Software Consortium (project number 024.003.037), Quantum
Delta NL (project number NGF.1582.22.030) and ENW-XL
grant (project number OCENW.XL21.XL21.122).
T.~R.~T. is supported by the Australian Research Council (FT220100359) and the U.S. Air Force Office of Scientific Research (FA2386-23-1-4062).

\section*{DATA AVAILABILITY}
The data and code that support the findings of this article are openly available \cite{github}.


\appendix

\section{Calculation of the photon absorption rate of the fictitious lasers for single EIT cooling}\label{App_A}

The photon absorption rate of the fictitious lasers in the steady state limit $R(\Delta)$ in Eq.\eqref{abs_rate} is calculated by solving the master equation
\begin{equation}\label{fokker_a}
    \dot{\rho} = ( \mathcal{L}_0 + \mathcal{L}_{-1}e^{-i\Delta t}  + \mathcal{L}_1e^{i\Delta t} ) \rho,
\end{equation}
employing the ansatz
\begin{equation}\label{ansatz}
    \rho(t) \approx \rho^{(0)} + \delta \rho^{(0)} + \delta \rho^{(+)}e^{-i\Delta t} + \delta \rho^{(-)}e^{i \Delta t},
\end{equation}
and using the matrix continued fractions method \cite{Athreya_time_Ham,fokker_planck} to calculate $\rho^{(0)} + \delta \rho^{(0)},\delta \rho^{(+)}$ and $\delta \rho^{(-)}$. The steady solution $\rho^{(0)} + \delta \rho^{(0)}$ can be obtained by solving 
\begin{equation}
    \left(\mathcal{L}_{-1}S_{1} + \mathcal{L}_0 + \mathcal{L}_{1}T_{-1}\right)(\rho^{(0)} + \delta \rho^{(0)}) = 0,
\end{equation}
where
\begin{flalign}
    S_1 &= -\left(\mathcal{L}_0 -i\Delta \right)^{-1}\mathcal{L}_1 \\ \nonumber
    T_{-1} &= -\left(\mathcal{L}_0 +i\Delta \right)^{-1}\mathcal{L}_{-1}, 
\end{flalign}
from which we can get $\delta \rho^{(+)}$ and $\delta \rho^{(-)}$ as
\begin{flalign}
    \delta \rho^{(+)} &= T_{-1}(\rho^{(0)} + \delta \rho^{(0)}) \\ \nonumber
    \delta \rho^{(-)} &= S_1(\rho^{(0)} + \delta \rho^{(0)}).
\end{flalign}
The rate of the excited state population $\dot{\rho}_{ee}$ is obtained by deriving the Bloch equations of Eq.\eqref{fokker_a} as
\begin{flalign}\label{bloch}
    \dot{\rho}_{ee} = -&\frac{i\Omega_1}{2}(\rho_{eg_{1}} - \rho_{g_{1}e})-\frac{i\Omega_2}{2}(\rho_{eg_{2}} - \rho_{g_{2}e}) \\ \nonumber
    + &\frac{\eta\Omega_1}{2}(\rho_{eg_{1}}e^{i\Delta t} + \rho_{g_{1}e}e^{-i\Delta t}) \\ \nonumber
    - &\frac{\eta\Omega_2}{2}(\rho_{eg_{2}}e^{i\Delta t} + \rho_{g_{2}e}e^{-i\Delta t}) - \Gamma \rho_{ee}.
\end{flalign}
Substituting Eq.\eqref{ansatz} into Eq.\eqref{bloch} and solving for the steady state $\dot{\rho}_{ee}$ = 0 we get
\begin{equation}
    \Gamma \rho_{ee}^{(0)} = -\frac{i\Omega_1}{2}(\rho^{(0)}_{eg_{1}} - \rho^{(0)}_{g_{1}e})-\frac{i\Omega_2}{2}(\rho^{(0)}_{eg_{2}} - \rho^{(0)}_{g_{2}e})
\end{equation}
since $\rho^{(0)}$ is the steady state solution under the influence of $\mathcal{L}_0$ and
\begin{flalign}\label{drho_bloch}
    \Gamma \delta\rho_{ee}^{(0)} = -&\frac{i\Omega_1}{2}(\delta\rho^{(0)}_{eg_{1}} - \delta\rho^{(0)}_{g_{1}e})\\ \nonumber
    -&\frac{i\Omega_2}{2}(\delta\rho^{(0)}_{eg_{2}} - \delta\rho^{(0)}_{g_{2}e}) \\ \nonumber
    + &\frac{\eta\Omega_1}{2}(\delta\rho_{eg_{1}}^{(+)} +\delta\rho_{g_{1}e}^{(-)}) \\ \nonumber
    - &\frac{\eta\Omega_2}{2}(\delta\rho_{eg_{2}}^{(+)} + \delta\rho_{g_{2}e}^{(-)}).
\end{flalign}
At the steady state, the rate of photon absorption is equal to the decay rate of the excited state population. Therefore, from Eq.\eqref{drho_bloch} we can conclude that the photon absorption rate due to the fictitious lasers at frequency $\Delta$ in the steady state limit is given by
\begin{flalign}
    R(\Delta) = \, &\frac{\eta\Omega_1}{2}(\delta\rho_{eg_{1}}^{(+)} + \delta\rho_{g_{1}e}^{(-)}) \\ \nonumber
    - &\frac{\eta\Omega_2}{2}(\delta\rho_{eg_{2}}^{(+)} + \delta\rho_{g_{2}e}^{(-)}).
\end{flalign}

\section{Double EIT}\label{double_eit}

\begin{figure}[htpb] 
\includegraphics[scale=1.0]{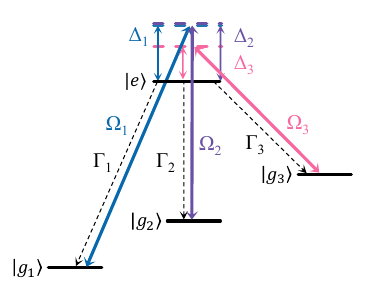}
\centering
\caption{Level scheme of double electromagnetically induced transparency cooling: Solid arrows connecting the $\ket{g_{k}} \rightarrow \ket{e}$ represent the three lasers with Rabi frequencies $\Omega_k$, detuned by $\Delta_k$. Dashed arrows represent decay from $\ket{e} \rightarrow \ket{g_k}$ with decay rate $\Gamma_k$.}
\label{double_levels}
\end{figure}

\begin{figure*}[t!]
\includegraphics[scale=1.0]{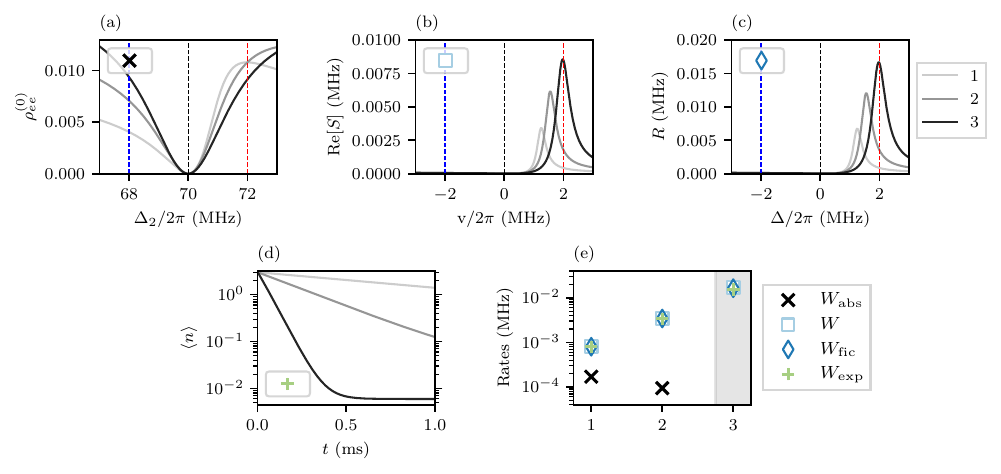}
\centering
\caption{Absorption profiles for double EIT cooling acquired from (a) the steady-state population of the rest ion's excited state, (b) the real part of the fluctuation spectrum of $V_1$, (c) the absorption rate of fictitious lasers and (d) the corresponding cooling dynamics for three scenarios:  (1) $\Omega_2/\Omega_1 = 0.5$, (2) $\Omega_2/\Omega_1 = 0.75$, and (3) $\Omega_2/\Omega_1 = 1$ for all $\Omega_1 = \Omega_3$. Dashed lines indicate the positions of the carrier (black), motion-removing (red), and motion-adding (blue) sideband transitions. (e) Cooling rates across these cases, where $W_{\text{abs}}$, $W$, $W_{\text{fic}}$ are obtained from spectra (a), (b), (c), respectively. $W_{\text{exp}}$ represents the numerically computed cooling rate which is used to benchmark our methods. The shaded region indicates negative $W_{\text{abs}}$ values. Note that  $W_{\text{abs}}$ has been multiplied by a factor of 1.33. Parameters: $\eta = 0.01$, $\omega_0/2\pi = 2$ MHz, $\Omega_1/2\pi = 16.5$ MHz, $\Delta_1/2\pi = \Delta_2/2\pi = 70$ MHz, $\Delta_2/2\pi  = 60$ MHz and $\Gamma/2\pi = $ 20 MHz with $\Gamma_1 = \Gamma_2 = \Gamma_3$}
\label{double_abs_all}
\end{figure*}

Double EIT cooling \cite{double_eit_theory1,double_eit_theory2,Semerikov2018,PhysRevLett.125.053001,Qiao2021} involves a closed four-level system consisting of three ground states $\ket{g_k}$ (k=1,2,3) connected to an excited state $\ket{e}$. The system is optically addressed by three lasers: two `drive' lasers couple the $\ket{g_1} \rightarrow \ket{e}$, $\ket{g_3} \rightarrow \ket{e}$ transitions with Rabi frequencies $\Omega_1$, $\Omega_3$, respectively and a counter-propagating `probe' laser drives the $\ket{g_2} \rightarrow \ket{e}$ transition with Rabi frequency $\Omega_2$. When the probe laser $\Delta_2$ detuning matches either $\Delta_1$ or $\Delta_3$, two dark states emerge, each corresponding to one of the resonance conditions. By selecting one of the two, we can take advantage of it similarly to single EIT cooling and bring our system near its motional ground state.

The Hamiltonian describing the system is given by
\begin{flalign}
    H' &= \omega_0 a^{\dagger}a + \Delta_1 \sigma_{g_1,g_1} + \Delta_2 \sigma_{g_2,g_2} + \Delta_3 \sigma_{g_3,g_3} \nonumber \\
    & -\frac{\Omega_1}{2} e^{ik_1 x}\sigma_{e,g_1} - \frac{\Omega_2}{2}e^{ik_2 x}\sigma_{e,g_2} - \frac{\Omega_3}{2}e^{ik_2 x}\sigma_{e,g_3} + \text{h.c}.
\end{flalign}

In first order of the LD regime, it can be expanded as
\begin{equation}
    H' \approx \omega_0 a^{\dagger}a + H'_R + V'_1(a + a^{\dagger})
\end{equation}
with
\begin{flalign}
    H'_R &= \Delta_1 \sigma_{g_1,g_1} + \Delta_2 \sigma_{g_2,g_2} + \Delta_3 \sigma_{g_3,g_3} \nonumber \\
    & -\frac{\Omega_1}{2} \sigma_{e,g_1} - \frac{\Omega_2}{2}\sigma_{e,g_2} - \frac{\Omega_3}{2}\sigma_{e,g_3} + \text{h.c}. \nonumber \\
    V'_1 &= -\frac{i\eta\Omega_1}{2}\sigma_{e,g_{1}} +\frac{i\eta\Omega_2}{2}\sigma_{e,g_{2}} -\frac{i\eta\Omega_3}{2}\sigma_{e,g_{3}} + \text{h.c.}
\end{flalign}
The spontaneous emission from the excited state to the ground is described by
\begin{equation}
    \mathcal{L'}\rho = \sum_{k}\Gamma_k \left[\sigma_{g_k,e}\rho\sigma_{e,g_k}-\frac{1}{2}\{\sigma_{e,e},\rho\}\right] 
\end{equation}

and the full dynamics are governed by
\begin{equation}
    \dot{\rho} = -i\left[ H', \rho\right] + \mathcal{L'}\rho 
\end{equation}

We can easily obtain the cooling rate $W$ by calculating the 
fluctuation spectrum Eq.\eqref{fluctuation spectrum} of operator $V'_1$ using
the superoperator $\mathcal{L}'_0: \mathcal{L}'_0\rho = -i[H'_R,\rho] + \mathcal{L'}\rho $ and the steady state of the rest ion $\widetilde{\rho}_{st}$, at the dark state condition. 

In terms of the fictitious lasers driving the sideband transitions, the system's Hamiltonian, in the rotating frame of the laser frequencies $\omega_k$, is described by
\begin{equation}
    H'_F = H'_R + H'_{-1} e^{-\Delta t} + H'_1e^{i\Delta t}  
\end{equation}
with
\begin{flalign}
    H'_{-1} &= -\frac{i\eta\Omega_1}{2}\sigma_{e,g_{1}} +\frac{i\eta\Omega_2}{2}\sigma_{e,g_{2}} -\frac{i\eta\Omega_3}{2}\sigma_{e,g_{3}}\\ \nonumber
    H'_1 &= (H'_{-1})^\dagger
    ,
\end{flalign}
where the fictitious laser frequencies $\widetilde{\omega}_k$ are chosen such that the detunings $\Delta = \widetilde{\omega}_k - \omega_k$ are identical for all $\ket{g_k} \rightarrow \ket{e}$ transitions. The system's dynamics are then governed by
\begin{equation}\label{fokker_d}
    \dot{\rho} = ( \mathcal{L}'_0 + \mathcal{L}'_{-1}e^{-i\Delta t} + \mathcal{L}'_1e^{i\Delta t}  ) \rho,
\end{equation}
where
\begin{flalign}
    \mathcal{L}'_{-1} \rho & = -i[H'_{-1},\rho], \nonumber \\
    \mathcal{L}'_1 \rho &= -i[H'_1,\rho].
\end{flalign}

Eq.\eqref{fokker_d} can be solved similarly as described in App.\ref{App_A} from which we obtain the absorption rate of the fictitious lasers at frequency $\Delta$ as
\begin{flalign}\label{abs_rate_double}
    R(\Delta) = \, -&\frac{\eta\Omega_1}{2}(\delta\rho_{eg_{1}}^{(+)} + \delta\rho_{g_{1}e}^{(-)}) \\ \nonumber
    + &\frac{\eta\Omega_2}{2}(\delta\rho_{eg_{2}}^{(+)} + \delta\rho_{g_{2}e}^{(-)}) \\ \nonumber
    - &\frac{\eta\Omega_3}{2}(\delta\rho_{eg_{3}}^{(+)} + \delta\rho_{g_{3}e}^{(-)}).
\end{flalign}
The cooling rate is caculated by
\begin{equation}
    W_{fic} = R(\omega_0)-R(-\omega_0).
\end{equation}

 Plotting the absorption profile of the rest ion, Fig.\ref{double_abs_all}(a), the fluctuation spectrum,  Fig.\ref{double_abs_all}(b), the absorption rate of the fictitious lasers, Fig.\ref{double_abs_all} (c) and the corresponding cooling dynamics, Fig.\ref{double_abs_all}(d), for a set of three parameters, we observe the success of the fluctuation spectrum and the absorption rate of the fictitious lasers to qualitatively predict the cooling rate according to their red-to-blue sideband absorption ratios. Similarly to single EIT, the absorption profile of the rest ion fails to make an accurate estimation for Rabi frequencies of a similar order, where optimal cooling is expected.Fig.\ref{double_abs_all}(e) reveals the quantitative agreement of $W$, $W_{fic}$ against $W_{exp}$ and the explicit failure of $W_{abs}$. The agreement/disagreement between $W$ and $W_{fic}$ can be used to distinguish between the weak/strong sideband coupling regime as previously established.

\section{Dynamics of the multilevel system}\label{dyn_multi}

The Hamiltonian describing the interaction of the cooling lasers on the $^{3}$D$_1$ $\leftrightarrow$ $^{3}$P$_0$ transition of $^{176}$Lu$^+$ for an arbitrary nuclear spin $I$ in the Schrödinger picture is

\begin{flalign}\label{Ham_multi_schro}
    \widetilde{H}_{R,S} &=  \sum_{m_{F_e}} g_{F_e} \mu_B m_{F_{e}} B \, \sigma_{F_e m_{F_{e}}, F_e m_{F_{e}}} \\ \nonumber
    &-\sum_{F_{g},m_{F_{g}}} \left(\omega_{F_e} - \omega_{F_g} - g_{F_g} \mu_B m_F B  \right)\sigma_{F_g m_{F_g}, F_g m_{F_g}} \\ \nonumber
    &-\sum_{\substack{F_g,m_{F_g} \\ m_{F_e},q}} \frac{\Omega^{F_g F_e \, q}_{m_{F_g} m_{F_e}}}{2}e^{-i \omega_{F_g F_e}t }\sigma_{F_e m_{F_e}, F_g m_{F_g}} + \text{h.c.}
\end{flalign}
where g$_\Phi$ are the Landé g factors (see App.\ref{lande}), $\tilde{\mu}_B$ = $\mu_B/\hbar$ with $\mu_B$ the Bohr magneton, $\sigma_{\Phi m_\Phi,\tilde{\Phi} m_{\tilde{\Phi}}}$ = $|\Phi, m_\Phi\rangle \langle \tilde{\Phi}, m_{\tilde{\Phi}}|$, $\omega_{F_gF_e}$ are the laser frequencies and $\omega_{F_e}$, $\omega_{F_g}$ are the frequencies of the $|F_e,m_{F_e} = 0\rangle, |F_g,m_{F_g} = 0\rangle$ levels respectively. $\Omega^{F_g F_e \, q}_{m_{F_g} m_{F_e}}$ are the Rabi frequencies between $|F_g,m_{F_g}\rangle \rightarrow |F_e,m_{F_e}\rangle $ determined by
\begin{equation}
    \Omega^{F_g F_e \, q}_{m_{F_g} m_{F_e}} = \frac{1}{\sqrt{2F_e + 1}}\bra{F_g m_{F_g} ; 1 q}\ket{F_e m_{F_e}} \Omega_{F_g F_e},
\end{equation}
where $\langle F_g m_{F_g} ; 1 q| F_e m_{F_e}\rangle$ are the Clebsch-Gordan coefficients and $\Omega_{F_gF_e}$ are the Rabi frequencies between $|F_g \rangle \rightarrow |F_e \rangle $. The Rabi frequencies are defined as 
\begin{equation}
    \Omega_{F_gF_e} = d_{F_gF_e}\mathcal{E}_{F_gF_e}
\end{equation}
with transition dipole moments 
\begin{equation}    d_{F_gF_e}=\bra{F_e}|e\mathbf{r}\cdot\mathbf{\hat{\epsilon}}_{q}|\ket{F_g},
\end{equation}
strengths of the electric fields of the lasers driving $|F_g \rangle \rightarrow |F_e \rangle $, $\mathcal{E}_{F_gF_e}$ and polarization vectors $\hat{\epsilon}_{q}$. The hyperfine transition dipole moments $d_{F_gF_e}$ are connected to the fine structure dipole moments $d_{J_gJ_e}$ as \cite{atomic}

\begin{flalign}
     d_{F_gF_e} &= (-1)^{F_g + J_e + I + 1}\\ \nonumber
    & \times \sqrt{(2F_e + 1)(2F_g +1)} 
    \begin{Bmatrix}
        J_e & F_e & I_g\\
        F_g & J_g & 1
    \end{Bmatrix} \,\, d_{J_gJ_e},
\end{flalign}
where the brackets \{\} indicate the 6j Wigner symbol.

The Hamiltonian Eq.\eqref{Ham_multi_schro} can be transformed into the interaction picture with $U=e^{iH_{0}t}$ where
\begin{equation}
    \widetilde{H}_{0} = \sum_{F_g,m_{F_g}} \omega_{F_gF_e}\sigma_{F_g m_{F_g},F_g m_{F_g}}.
\end{equation}

Then the Hamiltonian in the interaction picture becomes
\begin{flalign}\label{Ham_multi_inter}
    \widetilde{H}_{R,I} &=  \sum_{m_{F_e}} g_{F_e} \tilde{\mu}_B m_{F_e} B \, \sigma_{F_e m_{F_e}, F_e m_{F_e}} \nonumber \\
    &+ \sum_{F_g,m_{F_g}}\left( \Delta_{F_gF_e} + g_{F_g} \tilde{\mu}_B m_{F_g} B  \right)\sigma_{F_g m_{F_g}, F_g m_{F_g}} \nonumber \\
     &-\sum_{\substack{F_g,m_{F_g} \\ m_{F_e},q}} \frac{\Omega^{F_g F_e \, q}_{m_{F_g} m_{F_e}}}{2} \sigma_{F_e m_{F_e}, F_g m_{F_g}} + \text{h.c.}
\end{flalign}

Including motion, the Hamiltonian then becomes
\begin{flalign}\label{Ham_multi}
    \widetilde{H} &= \omega_0 a^{\dagger}a + \sum_{m_{F_e}} g_{F_e} \tilde{\mu}_B m_{F_e} B \, \sigma_{F_e m_{F_e}, F_e m_{F_e}} \nonumber \\
    &+ \sum_{F_g,m_{F_g}}\left( \Delta_{F_gF_e} + g_{F_g} \tilde{\mu}_B m_{F_g} B  \right)\sigma_{F_g m_{F_g}, F_g m_{F_g}} \nonumber \\
     &-\sum_{\substack{F_g,m_{F_g} \\ m_{F_e},q}} \frac{\Omega^{F_g F_e \, q}_{m_{F_g} m_{F_e}}}{2} e^{ik_{F_gF_e}x}\sigma_{F_e m_{F_e}, F_g m_{F_g}} + \text{h.c.}
\end{flalign}
where $\Delta_{F_gF_e} = \omega_{F_gF_e} -(\omega_{F_e}-\omega_{F_g})$ and $k_{F_gF_e}$ are the wavelengths of the lasers traveling along the x axis. Assuming that each of the laser beams has an equal superposition of $\sigma^{\pm}$
polarizations, the atom-laser part of the Hamiltonian should be multiplied by $1/\sqrt{2}$ and the relevant Rabi frequencies stated in the main text by $\sqrt{2}$ to replicate our results.

Similarly to the single EIT case, we can expand the Hamiltonian Eq. \eqref{Ham_multi} in the LD regime as
\begin{equation}\label{ld_eq_multi}
    \widetilde{H} \approx \omega_0 a^{\dagger}a + \widetilde{H}_R + \widetilde{V}_1(a + a^{\dagger}) 
\end{equation} 
where
\begin{flalign}
    \widetilde{H}_R &= \sum_{m_{F_e}} g_{F_e} \tilde{\mu}_B m_{F_e} B \, \sigma_{F_e m_{F_e}, F_e m_{F_e}} \nonumber \\
    &+ \sum_{F_g,m_{F_g}}\left( \Delta_{F_gF_e} + g_{F_g} \tilde{\mu}_B m_{F_g} B  \right)\sigma_{F_g m_{F_g}, F_g m_{F_g}} \nonumber \\
    &-\sum_{\substack{F_g,m_{F_g} \\ m_{F_e},q}} \frac{\Omega^{F_g F_e \, q}_{m_{F_g} m_{F_e}}}{2} \sigma_{F_e m_{F_e}, F_g m_{F_g}} + \text{h.c.} \nonumber \\
    \widetilde{V}_1 &= -\sum_{\substack{F_g,m_{F_g} \\ m_{F_e},q}} \frac{i\eta_{F_gF_e}\Omega^{F_g F_e \, q}_{m_{F_g} m_{F_e}}}{2} \sigma_{F_e m_{F_e}, F_g m_{F_g}} + \text{h.c.}
\end{flalign}
where $\eta_{(I-1)F_e} \approx \eta_{(I)F_e} \approx -\eta_{(I+1)F_e} \approx \eta$ = 0.044.

The spontaneous emission from $^3P_0$ to $^3D_1$ is described by 

\begin{widetext}
\begin{equation}\label{diss_multi}
    \mathcal{\widetilde{L}} \rho = \sum_{\substack{m_{F_e}\\F_g,m_{F_g}  }} \Gamma_{F_g m_{F_g},F_e m_{F_e}}\left[\sigma_{F_g m_{F_g},F_e m_{F_e}} \, \rho \,\sigma_{F_e m_{F_e,F_g m_{F_g}}}  
    - \frac{1}{2}\bigl\{ \sigma_{F_e m_{F_e},F_e m_{F_e}},\rho\bigl\}\right],
\end{equation}
\end{widetext}
where $\Gamma_{F_g m_{F_g},F_e m_{F_e}}$ are the decay rates between $\ket{F_e m_{F_e}} \rightarrow \ket{F_g m_{F_g}}$ (see App.\ref{decays}).
Finally, the dynamics of the system are governed by the master equation

\begin{equation}\label{dynamics_multi}
    \dot{\rho} = -i[\widetilde{H},\rho] + \mathcal{\widetilde{L}} \rho 
\end{equation}

\subsection{Landé g factors}\label{lande}
We use the Landé g factors calculated as if $I$ = 7 for all simulations independently of the nuclear spin $I$ under consideration. According to \cite{atomic} Landé g factors for hyperfine states obey the following equations:
\begin{align}
    g_F = 
    \begin{cases}
        g_J \frac{\text{F(F+1) +J(J+1) -I(I+1)}}{\text{2F(F+1)}} & \text{J} \neq 0 \\
        -g_I (m_e/m_p) & \text{J}=0,
    \end{cases}
\end{align}

\begin{equation}
    g_J = 1 + \frac{\text{J(J+1) + S(S+1) -L(L+1)}}{\text{2J(J+1)}}.
\end{equation}
Since that $J_e = 0$, we obtain $g_I$ by measurement of the nuclear magnetic moment $\mu_I = g_I \mu_N I$ of $^{176}$Lu$^+$ \cite{table,Lu_spec}. As a result, we get $g_I$ = 1.40431 or $g_I(m_e/m_p) \approx -2 \times 10^{-4}$. Finally, the g factors read

\begin{align}
    g_\Phi \approx
    \begin{cases}
        -2 \times 10^{-4} & F_e = I \\
        -0.0714 & F_g = I-1 \\
        0.0089 & F_g = I \\
        0.0625 & F_g = I+1 \\
    \end{cases}
\end{align}

\subsection{Decay rates} \label{decays}
From experimental measurements \cite{Lu_spec}, we know the decay rate between $^3P_0$ $\rightarrow$ $^3D_1$ of $^{176}$Lu$^+$, $\Gamma_{J_g,J_e}$ = 2$\pi$ $\times$ 2.45~MHz.
For our numerical simulations, we are interested in the decay rates between hyperfine states $\ket{F_e m_{F_e}}$ and $\ket{F_g m_{F_g}}$, $\Gamma_{F_g m_{F_g},F_e m_{F_e}}$. 
For fine structure states, we can write \cite{Einsteincoef,atomic}
\begin{flalign}\label{fine}
    \Gamma_{J_g,J_e} &= \frac{2e^2\omega_{J_g J_e}^3}{3\epsilon_0 h c^3} \frac{1}{(2J_e+ 1)}  \\ \nonumber
    &\times |\bra{n_e \, L_e \, S_e \, J_e}|\mathbf{r}|\ket{n_g\, L_g \, S_g \,J_g}|^2,
\end{flalign}
and for hyperfine structure states
\begin{flalign}\label{hyperfine}
    \Gamma_{F_g,F_e} &= \frac{2e^2\omega_{F_gF_e}^3}{3\epsilon_0 h c^3} \frac{1}{(2F_e + 1)} \\ \nonumber &\times|\bra{n_e\, L_e \, S_e \, J_e \, I_e \,F_e}|\mathbf{r}|\ket{n_g \, L_g \, S_g \, J_g \, I_g \, F_g }|^2,
\end{flalign}
where $e$ is the electron charge, $\omega_{eg}$ is the frequency of the $\ket{g} \rightarrow \ket{e}$ transition, $\epsilon_0$ is the vacuum permittivity,$h$ is the Planck constant, $c$ is the speed of light, $\bra{e}|e \mathbf{r}|\ket{g}$ is the dipole moment of the transition and we account for the degeneracy of the excited states $1/(2J_e(F_e) + 1)$. We can connect the two by the relation  \cite{atomic}:

\begin{flalign}\label{rel}
    &\bra{n_e \, L_e \, S_e \, J_e \, I_e \, F_e}|\mathbf{r}|\ket{n_g \, L_g \, S_g \, J_g \, I_g \,F_g} =  \nonumber \\
    &(-1)^{F_g + J_e + I_g + 1}\delta_{I_e,I_g} \times \bra{n_e \, L_e \, S_e \,J_e}|\mathbf{r}|\ket{n_g \, L_g \, S_g \, J_g} \nonumber \\
    & \times \sqrt{(2F_e + 1)(2F_g +1)} 
    \begin{Bmatrix}
        J_e & F_e & I_g\\
        F_g & J_g & 1
    \end{Bmatrix},
\end{flalign}
where $\delta_{I_e,I_g}$ is the Kronecker delta and the brackets \{\} indicate the 6j Wigner symbol. Using Eq. \ref{fine}, \ref{hyperfine}, \ref{rel}, we end up with
\begin{equation}\label{fgfe}
    \Gamma_{F_g,F_e} = (2J_e+1)(2F_g+1)
    \begin{Bmatrix}
        J_e & F_e & I_g\\
        F_g & J_g & 1
    \end{Bmatrix}^2
    \Gamma_{J_g,J_e}
    \left( \frac{\omega_{F_gF_e}}{\omega_{J_gJ_e}} \right) ^3.
\end{equation}
 We can drop the last term since $\left( \frac{\omega_{F_gF_e}}{\omega_{J_gJ_e}} \right) \simeq 1$. Adopting the normalization convention
\begin{flalign}\label{norm}
   &|\bra{n_e \, L_e \, S_e\,J_e\, I_e \,F_e}|\mathbf{r}|\ket{n_g \, L_g \, S_g \, J_g \, I_g \, F_g}|^2  =  \nonumber \\ &\sum_{m_{F_g},m_{F_e}}|\bra{n_e \, L_e \, S_e \, J_e \, I_e \, F_e \, m_{F_e}} \mathbf{r} \cdot \mathbf{\hat{\epsilon}}_{q}\ket{n_g \, L_g \,S_g \, J_g \, I_g \, F_g \, m_{F_g}}|^2
\end{flalign}
where q = $m_{F_e} - m_{F_g}$. Using Wigner-Eckart theorem, we obtain
\begin{flalign}
    &\bra{n_e \, L_e \, S_e \, J_e \, I_e \, F_e \, m_{F_e}}\mathbf{r} \cdot \mathbf{\hat{\epsilon}}_{q}\ket{ n_g \, L_g \, S_g \, J_g \, I_g \, F_g \, m_{F_g}} = \\ \nonumber
    &\frac{1}{\sqrt{2F_e +1}}\bra{F_g m_{F_g} ; 1 q}\ket{F_e m_{F_e}} \\ \nonumber
    & \times \bra{n_e \, L_e \,S_e \,J_e \,I_e \,F_e}|\mathbf{r}|\ket{n_g \, L_g \, S_g \, J_g \, I_g \, F_g} \\ \nonumber
    &= (-1)^{F_e-m_{F_e}} 
    \begin{pmatrix}
        F_e & 1 & F_g\\
        -m_{F_e} & q & m_{F_g} 
    \end{pmatrix}
    \\ \nonumber
    &\times \bra{n_e \, L_e \, S_e \, J_e \, I_e \, F_e}|\mathbf{r}|\ket{n_g \, L_g \, S_g \, J_g \, I_g \, F_g}
\end{flalign}
where  parentheses $\left(\right)$ indicate the 3j Wigner symbol. From Eq. \ref{fine}, \ref{fgfe}, \ref{norm}, \ref{wigner}
\begin{widetext}
\begin{equation}\label{wigner}
    \Gamma_{F_g,F_e} =\sum_{m_{F_g},m_{F_e}}\underbrace{\frac{2e^2\omega_{F_gF_e}^3}{3\epsilon_0 h c^3}\frac{1}{(2F_e+ 1)}  
    \begin{pmatrix}
        F_e & 1 & F_g\\
        -m_{F_e} & q & m_{F_g} 
    \end{pmatrix}^2
    |\bra{n_e \, L_e \, S_e \, J_e \, I_e \,F_e}|\mathbf{r}|\ket{n_g \,L_g \, S_g \,J_g \,I_g \,F_g}|^2
    }_{\Gamma_{F_g m_{F_g},F_e m_{F_e}}}
\end{equation}
\end{widetext}
where

\begin{flalign}\label{fmf}
    \Gamma_{F_g m_{F_g},F_e m_{F_e}} &= \begin{pmatrix}
        F_e & 1 & F_g\\
        -m_{F_e} & q &m_{F_g} 
    \end{pmatrix}^2 
    \Gamma_{F_g,F_e} \\ \nonumber
    & = \left(\frac{1}{2F_e +1}\right)\bra{F_g m_{F_g} ; 1 q}\ket{F_e m_{F_e}}^2 \Gamma_{F_g,F_e} 
\end{flalign}

Since we know $\Gamma_{J_g,J_e}$, we can easily calculate $\Gamma_{F_g,F_e}$ from Eq.\eqref{fgfe} and $\Gamma_{F_g m_{F_g},F_e m_{F_e}}$ from Eq.\eqref{fmf}. Lastly, we would like to comment on the normalization convention to avoid confusion. Although we used Eq.\eqref{norm} as our normalization convention, this choice is not unique. Another common normalization convention is
\begin{equation}
    |(F_e||\mathbf{r}|| F_g)|^2 = \sum_{m_{F_g}}|\bra{F_e m_{F_e}}\mathbf{r} \cdot \mathbf{\hat{\epsilon}}_{q}\ket{F_g m_{F_g}}|^2
\end{equation}

\subsection{Cooling optimization for multilevel systems}\label{cool_multi}
It's straightforward to calculate the fluctuation spectrum Eq.\eqref{fluctuation spectrum} of operator $\widetilde{V}_1$ using
the superoperator $\mathcal{\widetilde{L}}_0: \mathcal{\widetilde{L}}_0 \rho  = -i[\widetilde{H}_R,\rho] + \mathcal{\widetilde{L}}\rho $ and the steady state of the rest ion $\widetilde{\rho}_{st}$, both at the dark state condition, and therefore obtain the cooling rate $W$.

In terms of the fictitious lasers driving the sideband transitions, the system's Hamiltonian, in the rotating frame of the laser frequencies $\omega_k$, is described by
\begin{equation}
    \widetilde{H}_F = \widetilde{H}_R + \widetilde{H}_{-1} e^{-\Delta t} + \widetilde{H}_1e^{i\Delta t}  
\end{equation}
with
\begin{flalign}
\widetilde{H}_{-1} &= -\sum_{\substack{F_g,m_{F_g} \\ m_{F_e},q}} \frac{i\eta_{F_gF_e}\Omega^{F_g F_e \, q}_{m_{F_g} m_{F_e}}}{2} \sigma_{F_e m_{F_e}, F_g m_{F_g}}
     \\ \nonumber
     \widetilde{H}_1 &= (\widetilde{H}_{-1})^\dagger 
     ,
\end{flalign}
where the fictitious laser frequencies $\widetilde{\omega}_{F_gF_e}$ are chosen such that the detunings $\Delta = \widetilde{\omega}_{F_gF_e} - \omega_{F_gF_e}$ are identical for all $|F_g \rangle \rightarrow |F_e \rangle $ transitions. The system's dynamics are then governed by
\begin{equation}\label{fokker_m}
    \dot{\rho} = ( \mathcal{L}'_0 + \mathcal{L}'_{-1}e^{-i\Delta t} + \mathcal{L}'_1e^{i\Delta t}  ) \rho,
\end{equation}
where
\begin{flalign}
    \mathcal{L}'_{-1} \rho & = -i[H'_{-1},\rho]
    , \nonumber \\
    \mathcal{L}'_1 \rho &= -i[H'_1,\rho].
\end{flalign}

Eq.\eqref{fokker_m} is solved as established in App.\ref{App_A} from which we obtain the absorption rate of the fictitious lasers at frequency $\Delta$ 
\begin{flalign}
    R(\Delta) =  -\sum_{\substack{F_g,m_{F_g} \\ m_{F_e},q}}\left[\frac{\eta_{F_gF_e}\Omega^{F_g F_e \, q}_{m_{F_g} m_{F_e}}}{2} \right.\\ \nonumber
   \left. \times\left(\delta\rho_{F_e m_{F_e},F_g m_{F_g}}^{(+)} + \delta\rho_{F_g m_{F_g},F_e m_{F_e}}^{(-)} \right) \right]
\end{flalign}

where we now account for the absorption rates summed over all excited states. The cooling rate $W_{fic}$ is finally given by
\begin{equation}
    W_{fic} = R(\omega_0)-R(-\omega_0).
\end{equation}

\section{Imperfections in laser polarization}\label{App:polarization}

Assuming our laser propagates along the quantization axis, we introduce a small misalignment angle $\theta$. The polarization vector will then become 
\begin{equation}\label{pol_vec}
\hat{\epsilon} = -1/\sqrt{2}\, \text{cos}(\theta)\hat{\epsilon}_+ -\text{sin}(\theta)\hat{\epsilon}_0 + 1/\sqrt{2}\, \text{cos}(\theta)\hat{\epsilon}_-
\end{equation}
expressed in the polarization basis. This can be included in our Hamiltonian Eq.\eqref{Ham_multi} by multiplying the atom laser interaction with the appropriate coefficient from Eq.\eqref{pol_vec} for q=+1,0,-1.

Repeating the calculations of the main text for I=1, we find that the dark states are no longer independent of the Rabi frequencies as $\text{rank}[QHP_s]=2$ while $d_s$ remains 2 for both dark states formed for $\Delta_{11}/2\pi$ = 20.38, 20.5 MHz. Therefore, Eq.\eqref{dark_cond} is no longer satisfied. Decreasing the rank[$QHP_s$] requires unphysical conditions for the Rabi frequencies. For example, for the dark state formed for $\Delta_{11}/2\pi$ = 20.5 MHz, the Rabi frequenicies between $\ket{F_g=1,m_{F_g}=-1} \rightarrow \ket{F_e=1,m_{F_g}=-1}$, $\ket{F_g=1,m_{F_g}=-1} \rightarrow \ket{F_e=1,m_{F_g}=0}$, $\ket{F_g=2,m_{F_g}=1} \rightarrow \ket{F_e=1,m_{F_g}=1}$ and $\ket{F_g=2,m_{F_g}=1} \rightarrow \ket{F_e=1,m_{F_g}=0}$ should be switched off.

\begin{figure*}[t!]
\includegraphics[scale=1.0]{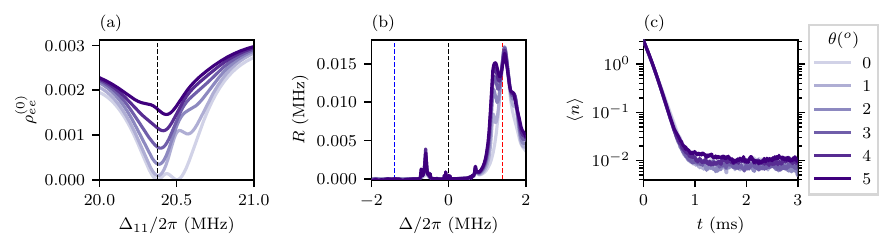}
\centering
\caption{$I$ = 1: Absorption profiles based on (a) the sum of the steady-state population of the rest ion's excited states and (b) the absorption rate of fictitious lasers for three scenarios for 5 misalignment angles $\theta$. Dashed lines indicate the positions of the carrier (black), motion removing (red), and motion adding (blue) sideband transitions. (c) The corresponding cooling dynamics averaged over 500 quantum trajectories. Parameters:$\eta = 0.044$, $\omega_0/2\pi = 1.4 $ MHz, B = 5 G, $\Omega_{01}/2\pi = \Omega_{21}/2\pi$ = $\sqrt{2} \times$23.5 MHz, $\Omega_{11}/2\pi$ = $\sqrt{2}\times$10 MHz ($\Omega_{F_g F_e}$), $\Delta_{01}/2\pi = 10$ MHz, $\Delta_{11}/2\pi = 20.38$ MHz, $\Delta_{21}/2\pi = 20$ MHz ($\Delta_{F_g F_e}$) and $\Gamma_{J_gJ_e}/2\pi = 2.45 $ MHz.}
\label{I1_abs_pol}
\end{figure*}

Fig.\ref{I1_abs_pol}(a) shows tha absorption profile of the rest ion. The dark state formed at $\Delta_{11}/2\pi = 20.5$ MHz exhibits strong sensitivity to polarization imperfections, while the dark state at $\Delta_{11}/2\pi = 20.38$ MHz is significantly more robust. Consequently, $\Delta_{11}/2\pi = 20.38$ MHz represents a more favorable choice for an assumed experimental implementation of the toy model with I=1. Non-zero $\rho_{ee}^{(0)}$ population gives rise to finite absorption on the carrier transition. This alone does not permit a direct prediction of the cooling rates. To address this, we calculate the absorption of the fictitious lasers Fig.\ref{I1_abs_pol}(b). The results indicate that the absorption on the motion-adding sideband remains suppressed, while on the motion-adding sideband it remains enhanced at all angles of misalignment $\theta$. This suggests that cooling should still occur, however with a higher final mean phonon number due to the finite absorption on the carrier transition. This prediction is in agreement with the calculated cooling dynamics in Fig.\ref{I1_abs_pol}(c).

\begin{figure*}[t!]
\includegraphics[scale=1.0]{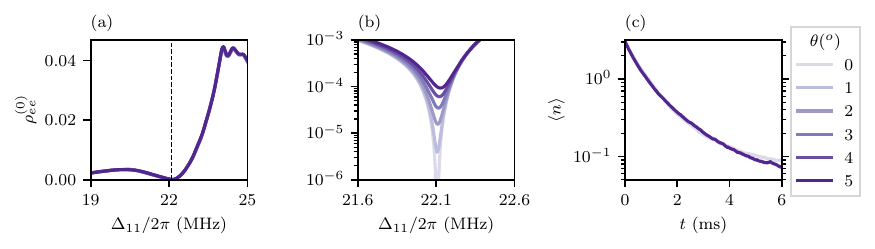}
\centering
\caption{$I$ = 7: (a) Absorption profile based on the sum of the steady-state population of the rest ion's excited states and (b) its magnified view for 5 misalignment angles $\theta$. The dashed black line indicates the positions of the carrier sideband. (c) The corresponding cooling dynamics averaged over 500 quantum trajectories. Parameters: $\eta = 0.044$, $\omega_0/2\pi = 1.4 $ MHz, B = 4 G, $\Omega_{67}/2\pi = \Omega_{87}/2\pi$ = $\sqrt{2} \times$70 MHz, $\Omega_{77}/2\pi$ = $\sqrt{2}\times$19 MHz ($\Omega_{F_g F_e}$), $\Delta_{67}/2\pi = 10$ MHz, $\Delta_{77}/2\pi = 22.1$ MHz, $\Delta_{87}/2\pi = 20$ MHz ($\Delta_{F_g F_e}$) and $\Gamma_{J_gJ_e}/2\pi = 2.45 $ MHz.}
\label{I7_abs_pol}
\end{figure*}

For the model with I=7, polarization imperfections also reduce the rank[$QHP_s$] from 2 to 1 while $d_s$ remains 2. However, the dark state formed at $\Delta_{77}/2\pi = $
22.1 MHz is quite robust as shown in Fig.\ref{I7_abs_pol}, and consequently the cooling dynamics are not significantly affected. Due to computational expense, we only compute the cooling dynamics for $\theta$ = 0$\degree$ and 5$\degree$. The spectrum of the fictitious lasers is omitted, as it does not provide relevant information in the ultra-strong coupling regime, as explained in the main text. 

For both I=1 and I=7, the recoil term in the spontaneous emission \cite{Cirac_Zoller_cooling,Molmer:93} doesn't drastically affect the dynamics due to the non-zero absorption on the carrier transition and is neglected in the numerical calculations. 

\bibliography{draft_2}


\end{document}